\documentclass[%
 aip,
 amsmath,amssymb,
 reprint,%
]{revtex4-1}

\usepackage{graphicx}
\usepackage{dcolumn}
\usepackage{bm}

\usepackage[utf8]{inputenc}
\usepackage[T1]{fontenc}
\usepackage{mathptmx}
\usepackage{etoolbox}

\makeatletter
\def\@email#1#2{%
 \endgroup
 \patchcmd{\titleblock@produce}
  {\frontmatter@RRAPformat}
  {\frontmatter@RRAPformat{\produce@RRAP{*#1\href{mailto:#2}{#2}}}\frontmatter@RRAPformat}
  {}{}
}%
\makeatother
\begin{document}

\preprint{AIP/123-QED}

\title[Penetrative convection in nocturnal atmospheric boundary layer and radiation fog]{Penetrative convection in nocturnal atmospheric boundary layer and  radiation fog }
\author{Shaurya Kaushal}
\author{D.K. Singh}
\author{K.R. Sreenivas*}%
 \email{krs@jncasr.ac.in}
\affiliation{ 
Engineering Mechanics Unit, Jawaharlal Nehru Centre for Advanced Scientific Research, Jakkur, Bangalore  560064, India
}%

\date{\today}

\begin{abstract}
After the sunset, under calm and clear sky conditions, aerosol laden surface air-layer, cools radiatively \cite{mukund,singh} to the upper
atmosphere. Predominant effect of the radiative cooling on the vertical temperature profile extends to several hundred meters
from the surface. This results in the development of a stable, nocturnal inversion layer. However, ground surface, owing to its
higher thermal inertia, lags in the cooling process. Due to this about a meter thick air layer just above the ground can be $(2-6)^\circ$C
cooler than the ground \cite{ramdas, blay2015}. Thus, at the surface an unstable convective layer is present, which is capped by a stable inversion layer
that extends up to several hundred meters. This configuration involving a convective mixed layer topped by a stably
stratified inversion layer is a classic case of penetrative convection \cite{townsend,adrian}. Micro-meteorological phenomenon at the surface,
such as occurrence of fog, is determined by temperature profile, heat and moisture transport from the ground. Here, we
present a computational study of the model penetrative convection, formed due to radiative cooling, in the nocturnal atmospheric
boundary layer. 
\end{abstract}

\maketitle

\section{\label{sec:level1}Introduction}
Penetrative convection, as  defined by one of the earlier studies \cite{veronis}, involves a convective mixed layer topped by a stably stratified non-turbulent layer. Associated  dynamics of penetrative motion of the fluid from the mixed layer into stable layer across  the interface, results in entrainment and growth of the mixed layer. This phenomena finds relevance in a wide variety of geophysical and astrophysical problems. The growth of atmospheric daytime boundary layer \cite{driedonks},
motions in the outer stable regions of the sun \cite{leighton}, 
the deepening of upper-ocean mixed layer,
dynamics of cumulus clouds \cite{simpson},
turbulent convection in water over an ice surface \cite{townsend, adrian} and 
development of radiation-fog inversion layer
are some of the examples.

In all such systems, the two major areas of interest are: 
$1)$ the convectively driven `mixed layer' and 
$2)$ dynamics and fluxes through interfacial region sandwiched between turbulent mixed layer and the non-turbulent stably stratified region, often referred to as the `entrainment zone'\cite{deardorff1980}. 
In order to better understand the formation and subsequent deepening of the convectively driven mixed layer, experimental studies \cite{willis1974, kumar} on the structure of turbulence have successfully established the integral length scale and the horizontal velocity scale in the bulk. The length scale is nothing but the depth of the mixed layer (h) and the convection velocity ($U^*$) is defined as\cite{deardorff1980}, 
\begin{equation}
U^* = \left[ \frac{g\beta Q_b h}{\rho C_p} \right]^{1/3},
\end{equation}
where, $g$ is the gravitational constant, $\beta$ is the coefficient of thermal expansion, $\rho$ is the reference density, $C_p$ is the specific heat for fluid at constant pressure and $Q_b$ is the bottom heat flux.
Experimental studies indicate that the thickness of the entrainment zone ($Z_p$) could typically reach upto 25 percent of that of the mixed layer \cite{deardorff1980}.
These studies \cite{sreenivas, deardorff, deardorff1980} emphasize on the importance of entrainment zone physics in understanding phenomena related to the growth/depletion of the mixed layer.  
The normalized entrainment rate $(U_e/U^*)$, has been shown to be a function of the Richardson number $(Ri)$, with a pre-factor $c$ and an exponent of $n$, 
\begin{equation}
\frac{U_e}{U^*} = c \> \left( Ri^{-n} \right),
\end{equation}
where, Richardson number is a widely used non-dimensional parameter
that is used to predict the occurrence of fluid turbulence.
The definition of $Ri$ in the context of this study is,
\begin{equation}
Ri = \frac{g \left(\Delta \rho \right) h}{\rho (U^*)^2} 
\end{equation}
$\Delta \rho$ is the density jump across the interface.

Most of the laboratory experiments on penetrative convection  \cite{deardorff,kumar,fernando,sreenivas}, conducted to establish the aforementioned equations, consist of a stably stratified layer of water  and the convection in the system is driven by higher
temperature/heat flux at the lower boundary. These configurations allow for a robust and simple experimental controlled environment, but tend to leave out a certain class of penetrative convection problems which find application in atmospheric boundary layers.
 
The focus of this paper is to study penetrative convection in a class of problems where the system is not driven by a heated boundary, but, by a spatially varying volumetric source/sink term. 
An aerosol laden atmosphere is a relevant example of such a system, where radiative cooling caused by aerosols, results in some interesting atmospheric phenomena. We specifically focus on the peculiar nature of the nocturnal boundary layer that  dates back to
Ramdas and co-workers in the 1930’s \cite{ramdas}, where a Lifted Minimum
Temperature(LTM) profile was first observed. The origin of which has been demonstrated  to be related to radiative flux divergence arising due to the presence of aerosols in the nocturnal atmospheric boundary layer \cite{mukund, singh}. The Ramdas paradox is the occurrence of a counter intuitive temperature minimum, a few tens of centimeters above the ground, on calm and clear nights. After nightfall, under calm and clear sky conditions, atmospheric surface layer close to the ground is shown to cool by aerosol radiation, whereas, the ground lags behind due to its higher thermal inertia \cite{singh,blay2015}. This results in the formation of a convectively driven mixed layer capped above by a stable inversion layer. Under certain environmental conditions, the radiative cooling in the mixed layer causes the air to reach saturation and eventually leads to fog formation. This radiation fog formed near the surface of the ground, thickens, as the air continues to cool and the deepens (overnight), as the stable layer above the fog gets entrained.

\begin{figure}
	\includegraphics[scale=0.45]{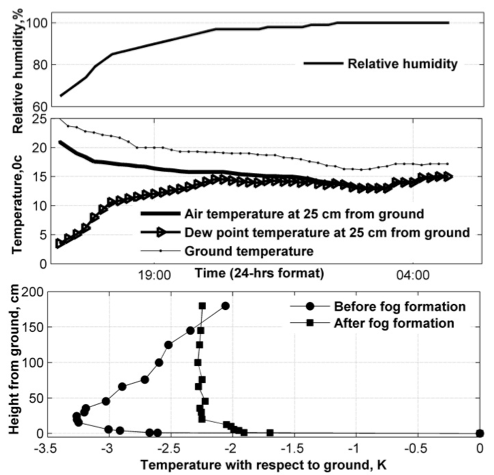}
	\caption{\label{dew} Relative humidity, air temperature, and dew-point temperature at a height of 25 cm above the ground, and ground temperature as a function of time during the night. The vertical temperature profiles above the ground before and after the formation of fog.}
\end{figure}

In a recent experimental study by Mukund et al.\cite{mukund}, major emphasis has been laid on the role of radiative cooling by aerosols in the formulation of the nocturnal boundary layer. It has been established that the inclusion of the radiative flux term in the model, is crucial to obtaining the correct temperature profiles \cite{singh}, as seen in Fig.\ref{dew}. 
Fig.\ref{dew} depicts the importance of knowing local temperature and dew-point temperature for predicting the onset of fog. It also depicts how fog (liquid aerosol particles)
could modify vertical temperature profile through modifying radiation cooling.
Experiments carried out by Hutchison and Richards \cite{hutchison} lay emphasis on the same by studying the effect of radiation on the onset of convection, using carbon dioxide as the participating medium.
Recent studies on the Martian atmosphere have also reported the presence of aerosols to be a major cause for fog formation \cite{mars}.

In this work, we use the established theoretical understanding of nocturnal boundary layer and radiation fog, to setup a two dimensional simulation domain of $1$m vertical height, with the intent of capturing the LTM, mixed layer and entrainment zone dynamics and subsequently, quantitatively analyze the relationship between the entrainment rate and Richardson number. Dynamics in this layer is typical and has an impact on the occurrence and deepening of the fog layer developed due to radiative cooling \cite{mukund}. The correct aerosol number density is important for correctly estimating the radiative flux term. Hence, field and laboratory experiments (see details \cite{dhiraj_thesis}) were also carried out in order to estimate the nocturnal aerosol density at relevant heights close to earth's surface.    

Section \ref{gov_eq} of this paper describes the governing equations for the system and also sheds light on field and laboratory experiments carried out to correctly determine the aerosol density, a few decimeters above the surface of the earth. Section \ref{comp_set} describes the computational setup for running the simulations, highlighting the initial and boundary conditions used in order to closely replicate nocturnal atmospheric conditions. In reality, atmospheric conditions such as aerosol density are seldom consistent and hence, we run the simulation for test cases that span three orders of magnitude in particle number density. We present the mean temperature, density and heat flux profiles in Section \ref{res1}, followed an analysis of the mixed layer and the entrainment zone in Section \ref{res2}. 


\section{\label{gov_eq}Governing equations}
\begin{figure}
	\includegraphics[scale=0.45]{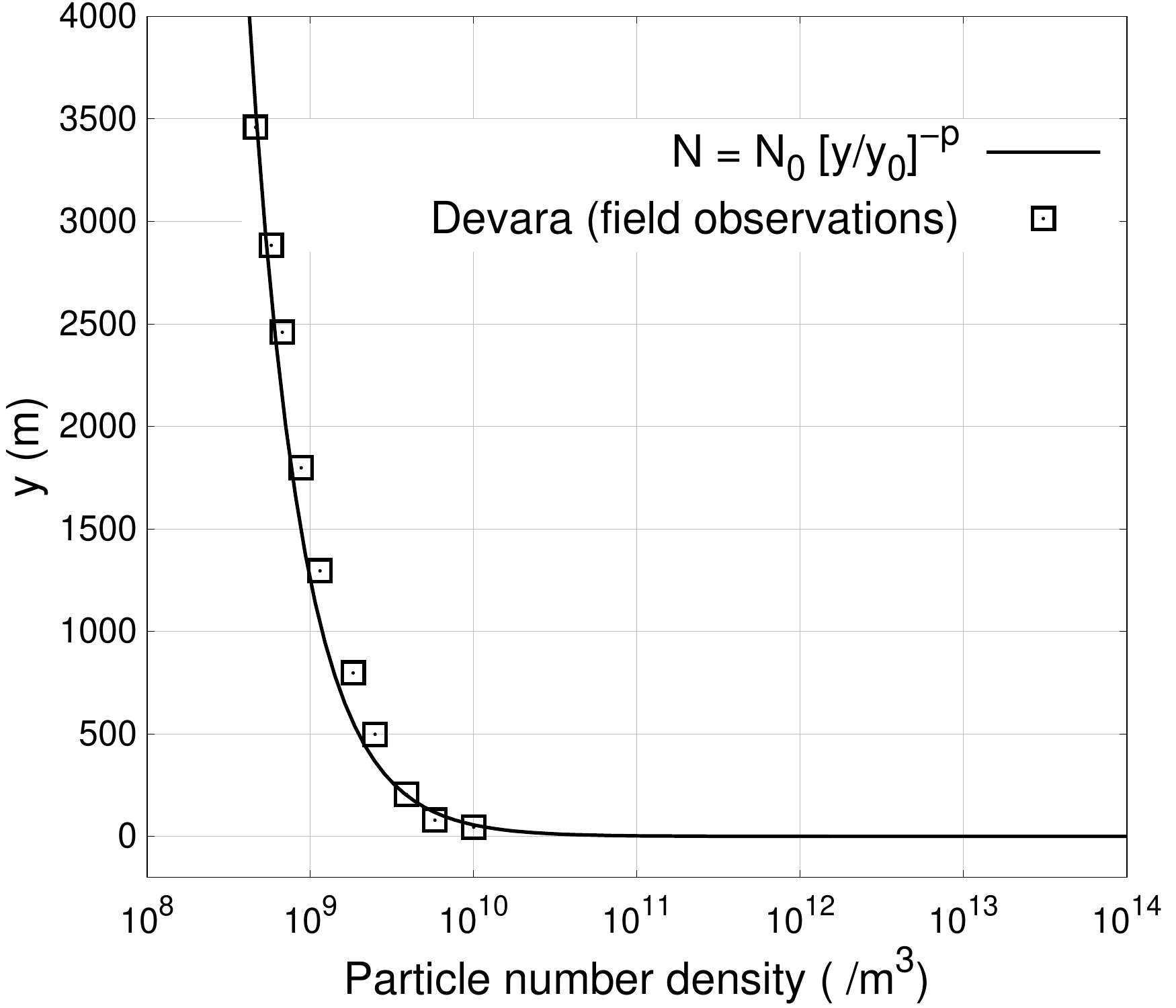}
	\caption{\label{N_far} Vertical variation in the number density in a stable, nocturnal boundary layer
		reported by Devara\cite{devara1993,devara1997} along with a Rouse
		profile fit to the data, where $N_0=10840$ x$10^6$, $y_0=50$ and $p=0.74$}
\end{figure}

\begin{figure}
	\includegraphics[scale=0.45]{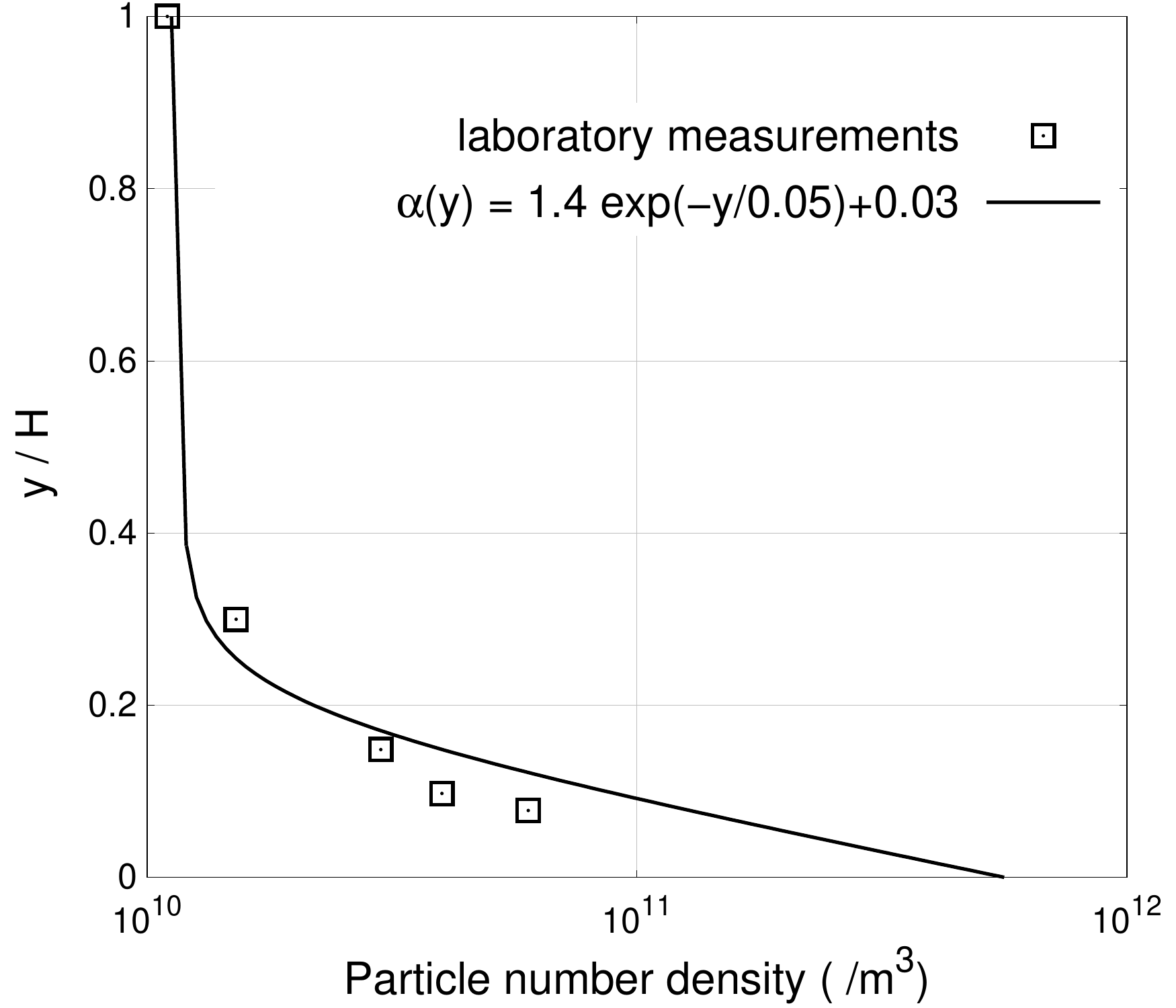}
	\caption{\label{N_close} Vertical number density, in the laboratory test section plotted with respect to
		normalized length scale (y/H, with H=$10$ cm) and $\alpha(y)$ is the non-dimensional absorption coefficient in the test section \cite{dhiraj_thesis}}
\end{figure}

 The governing equations for the system are conservation of mass, momentum and energy. The system is two dimensional for all purposes in the scope of this paper, with `$u$' being the velocity component in horizontal ($x$) direction and `$v$' being the velocity component in the vertical ($y$) direction. The density, temperature and dynamic viscosity  of air is given by $\rho$, $T$ and $\mu$.   
\begin{eqnarray}
\frac{d \rho}{dt} + \frac{\partial (\rho u)}{\partial x} + \frac{\partial (\rho v)}{\partial y} &&= 0 ,
\\
\rho \left( \frac{du}{d t} + u \frac{\partial u}{\partial x} + v \frac{\partial u}{\partial y}\right) &&=   \mu \left( \frac{\partial^2}{\partial x^2} + \frac{\partial^2}{\partial y^2} \right)  u ,
\\
\rho \left( \frac{dv}{d t} + u \frac{\partial v}{\partial x} + v \frac{\partial v}{\partial y}\right) &&=  \mu \left( \frac{\partial^2}{\partial x^2} + \frac{\partial^2}{\partial y^2} \right)  v - \rho g ,
\\
\rho C_p \left( \frac{\partial T}{\partial t} + u\frac{\partial T}{\partial x} + v \frac{\partial T}{\partial y} \right) &&= k \left( \frac{\partial^2}{\partial x^2} + \frac{\partial^2}{\partial y^2} \right) T + q''' .
\label{governing}
\end{eqnarray}
where, the $g$ is the gravitational constant, $C_p$ is the specific heat at constant pressure, $k$ is the thermal conductivity and $q'''$ is the radiative flux divergence term. 
The total radiative flux divergence $(q''')$ is modelled by multiplying the local aerosol number density ($N$) with the net radiative forcing from a single aerosol particle ($Q_R$, refer Eq.\ref{QR}),  
\begin{equation}
q''' =  N \times Q_R.
\label{q'''}
\end{equation}
The non-uniform vertical distribution of aerosols, as indicated in Fig.\ref{N_far}, is the primary
reason for the observed hyper-cooling being close to the ground. Such a steep variation in
number density is common for particulate matter suspended in a fluid \cite{soulsby, nielsen} and is often modelled using a Rouse-profile\cite{devara1993, devara1997}. 
It is worth noting, 
that in field experiments, the typical height for measurements using lidar, starts from $50$m and extends up to $4$km. In the context of studies revolving around NBL and LTM, the height at which relevant
dynamics takes place, we need information within few meters close to
ground. Hence, aerosol number densities as a function of height, much
closer to the ground, were estimated in a laboratory test section\cite{dhiraj_thesis}.The minimum height from
bottom boundary where images are taken is $8$mm, as shown in Fig.\ref{N_close}.  

The particle number density as seen in Fig.\ref{N_close}, is given by, 
\begin{equation}
N = \frac{1}{A} \left[1.4 \exp \left(-\frac{y}{0.05}\right) + 0.03  \right],
\label{N}
\end{equation}
and the radiatve flux balance on a single aerosol particle is,
\begin{equation}
Q_R = - \frac{\sigma A}{2} \epsilon_p \left( \epsilon_g T_g^4 + T_{sky}^4(2-\epsilon_g) - 2 T_p^4\right),
\label{QR}
\end{equation}
 where, $\sigma$ denotes the Stefan Boltzmann constant and $A$ is the radiative area of spherical aerosol particle. We consider a simplistic case of uniform diameter aerosol particles with a diameter $d = 1\mu$m, so the surface area of the sphere is simply $A = \pi d^2 $(note here that, it is the product of number density N and
 surface area, that will become important). In Eq.\ref{QR}, $T_p$ denotes the temperture of the aerosol particle at a given location, $T_g$ is the ground temperature, $T_{sky}$ is the sky temperature, $\epsilon_g$ is the ground emissivity and $\epsilon_p$ is an area-averaged aerosol emissivity. Corresponding values for temperatures and emmisivities are taken from laboratory experiments \cite{mukund}, with $T_g = 298 K$, $T_{sky} = 273 K$ and $\epsilon_g = \epsilon_p = 0.9$.  

The dynamics of aerosol particles is governed by a simple advection-diffusion equation (see Eq.~(\ref{conc}).) where the non-dimensional concentration is given by $C$, such that,  
\begin{equation}
C =  \frac{N- N_t}{N_b - N_t}
\end{equation}
where $N_t$ and $N_b$ are the number density of particles at the top and bottom of the computational domain evaluated using Eq.\ref{N}. The advection-diffusion for the aerosol is given by,
\begin{equation}
\frac{dC}{dt} + \left(u\frac{d}{dx} + v \frac{d}{dy} \right)C = \delta \left( \frac{\partial^2}{\partial x^2} + \frac{\partial^2}{\partial y^2} \right)C 
\label{conc}
\end{equation}
where, $\delta$ is the diffusivity.

\section{\label{comp_set}Computational setup}
\begin{figure}
	\includegraphics[scale=0.3]{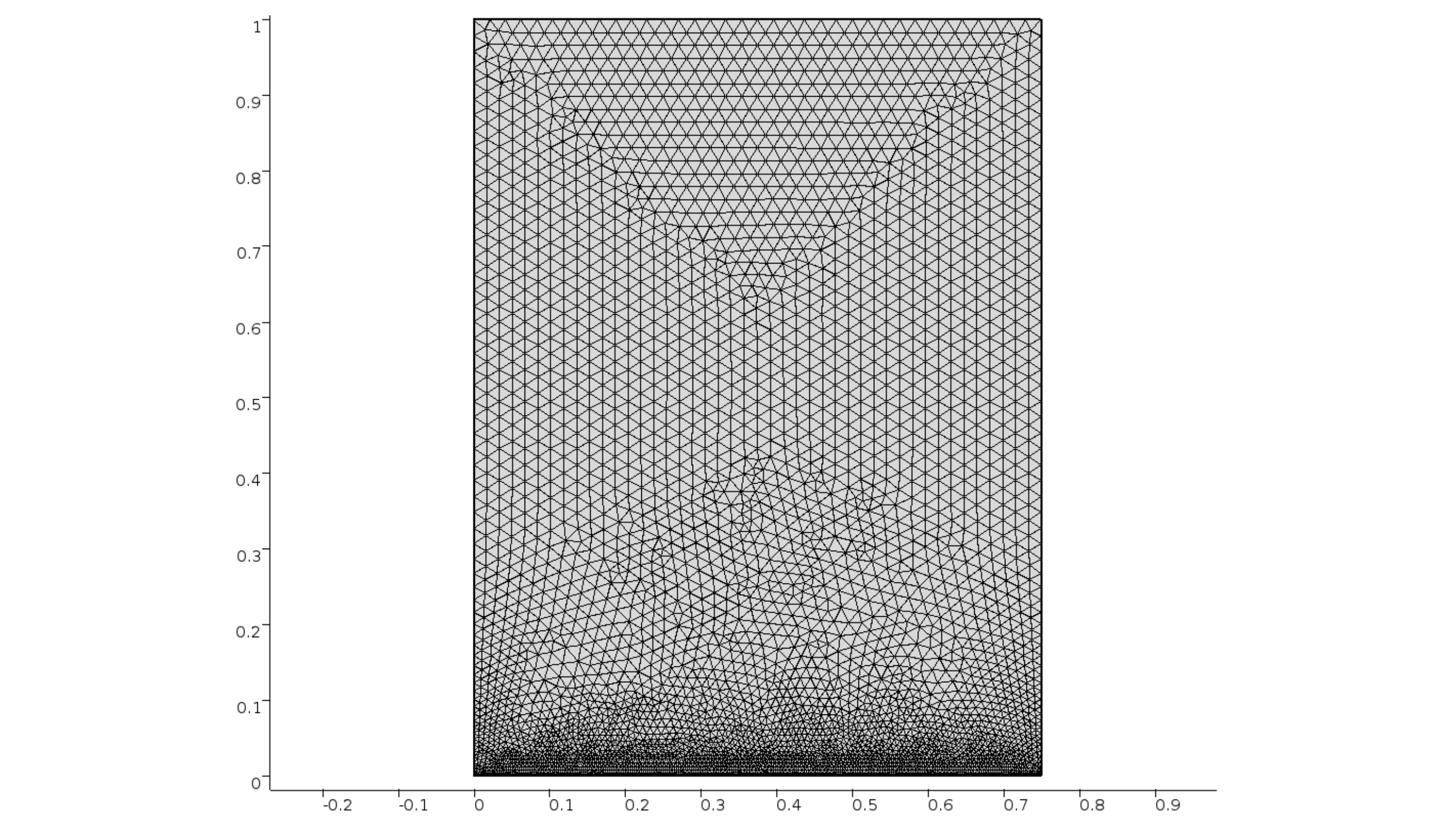}
	\caption{\label{grid} Triangular grid used for computation}
\end{figure}

The computational setup is a 2D rectangular domain, $1\text{m}$ in height and $0.75\text{m}$ in width as shown in Fig.~\ref{grid}. The height of the domain is large enough to incorporate the lifted minimum temperature and the mixed layer turbulent dynamics, which is observed a few decimeters above the surface of the ground. The side walls of the domain are kept periodic for mass, momentum and energy in an attempt to model a relatively large section of horizontal ground. Fog can usually span horizontal length scales of few hundred meters and hence, a periodic boundary for the side walls is appropriate for all practical purposes in scope of the problem.
The bottom boundary in non-penetrable, non-slip and fixed at a temperature $T_g$. For the simulations presented in this paper, we have taken it to be $T_g= 295.15$K. The top boundary is intended to mimic the night sky at $1\text{m}$ height and an open boundary condition is applied along with a constant heat flux, $ q_t(=0.1W/m^2)$.
Unlike experiments, where a Dirichlet boundary condition is applied on the top boundary, numerical simulations of this kind facilitate us to use a more realistic radiative flux boundary condition, mimicking radiative cooling and the lapse rate typically observed at $1$m height in the field experiments \cite{mukund}.
The simulation is initialised with a zero velocity field across the domain and a linear initial temperature profile given by, $T_0 = \Delta T \> y + T_g$, where, $\Delta T = 2K$. 

Arguably, the most important step in achieving the LTM profile is modelling the radiative cooling term in the energy equation. As show in Eq.\ref{q'''}, aerosols are the source of the radiative cooling and flux term is calculated as a product of the number of aerosol particles and the radiative flux from each particle. The initial number density of aerosols as a function of height is taken to be \cite{singh}, 
\begin{equation}
N_0 = N_p \times \frac{1}{\pi}\left[ 1.4\exp \left(-\frac{y}{0.05} \right)+0.03\right] \times 10^{12},
\end{equation}
as observed in experiments\cite{singh}, where the number density of aerosol particles was measured as close to $8$mm from the ground under clear and calm night conditions. 
$N_p$ is a parameter used to vary the aerosol density and hence the radiative cooling. We run the test problem for a set of three aerosol number densities (see Table\ref{table1}), labelled as Case a), b) and c), with the intent of covering a wide variety of atmospheric conditions that can lead to different orders of magnitude of particle number density. The initial concentration $C_0$, for all the cases follows a linear profile with the bottom boundary value of $1$ and a top boundary condition of $0$. The simulation is run for a total time of $10$ minutes with $\Delta t = 0.01$ seconds on COMSOL with an extremely fine grid size .
\begin{table}
	\caption{\label{table1} }
	\begin{ruledtabular}
		\begin{tabular}{cccc}
			$q_{t}= 0.1 \> W/m^2$&$N_p$  &$N_b \> (/m^3)$ & $N_t \> (/m^3)$\\ \hline \\
			 Case a)             & $0.1$ & $4.55 \times 10^{10}$ & $9.54 \times 10^{8}$\\
			 Case b)             & $1$   & $4.55 \times 10^{11}$ & $9.54 \times 10^{9}$\\
			 Case c)             & $10$  & $4.55 \times 10^{12}$ & $9.54 \times 10^{10}$\\
		\end{tabular}
	\end{ruledtabular}
\end{table}

\section{Results}
\subsection{\label{res1} Mean temperature, density and flux profiles}

\begin{figure*}
	\centering
	\begin{tabular}[b]{c}
		\includegraphics[width=.28\linewidth]{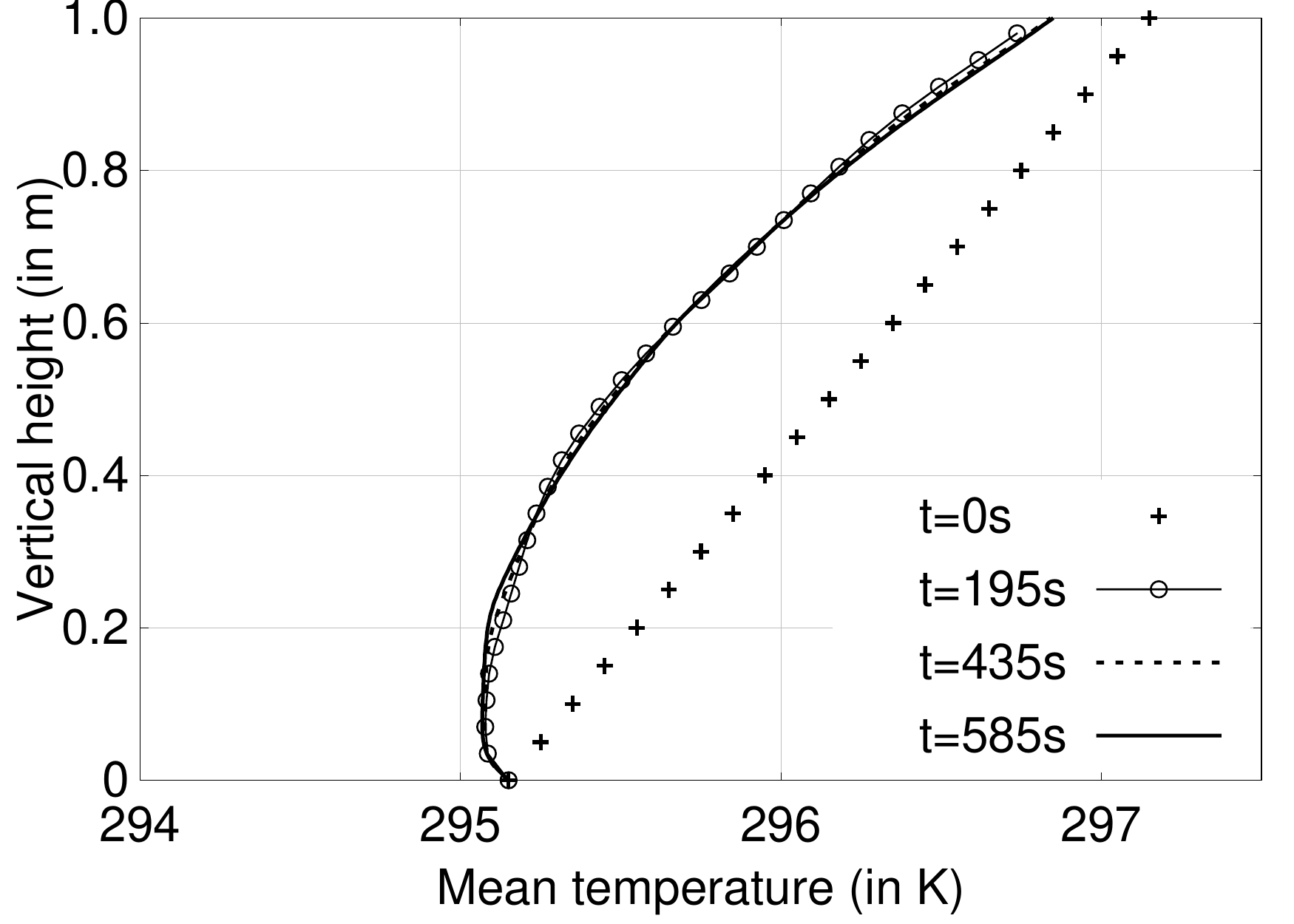} \\
		\small Case a) $\hspace{0.3cm} N_p = 0.1$
	\end{tabular} \qquad
	\begin{tabular}[b]{c}
		\includegraphics[width=.28\linewidth]{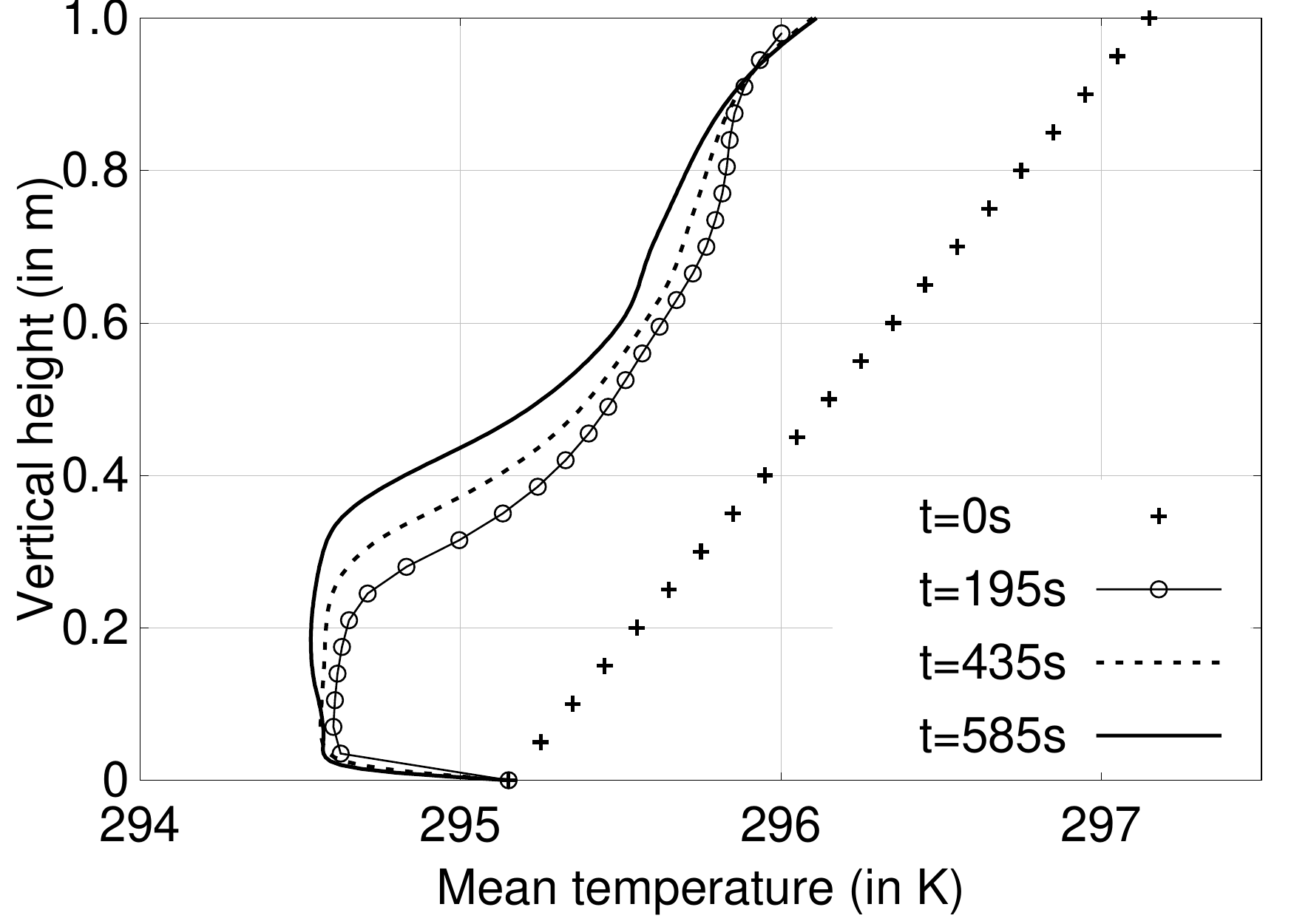} \\
		\small Case b) $\hspace{0.3cm} N_p = 1$
	\end{tabular} \qquad
	\begin{tabular}[b]{c}
		\includegraphics[width=.28\linewidth]{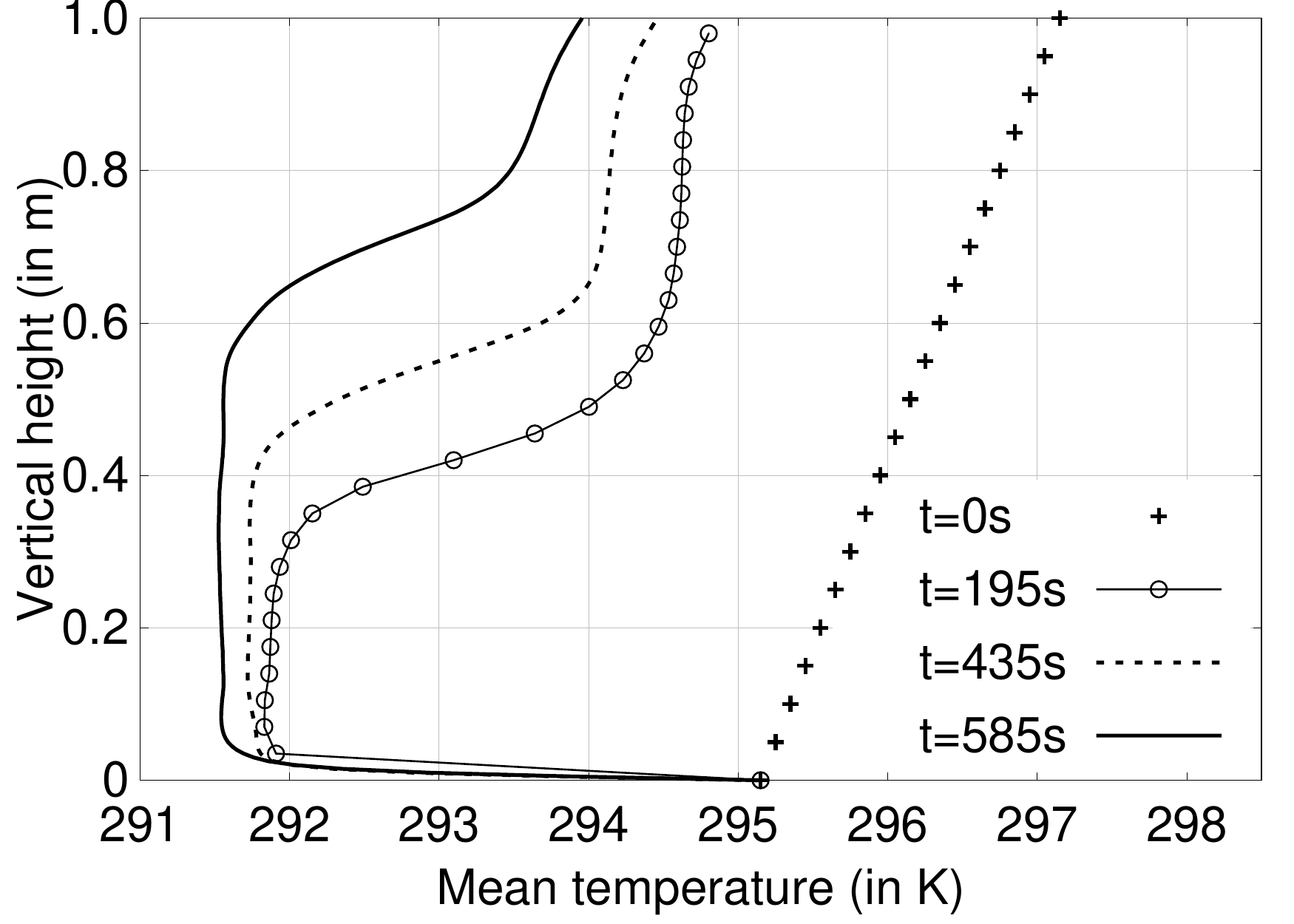} \\
		\small Case c) $\hspace{0.3cm} N_p = 10$
	\end{tabular}
	\caption{\label{temp}Mean temperature profiles as a function of vertical height, plotted for four timestamps $t=0,195,435,585$ seconds. The radiative cooling caused by the aerosols leads to the formation of LTM profiles, with growing mixed layers in Case b) and c) and a stagnant mixed layer in Case a).}
\end{figure*}

\begin{figure*}
	\centering
	\begin{tabular}[b]{c}
		\includegraphics[width=.28\linewidth]{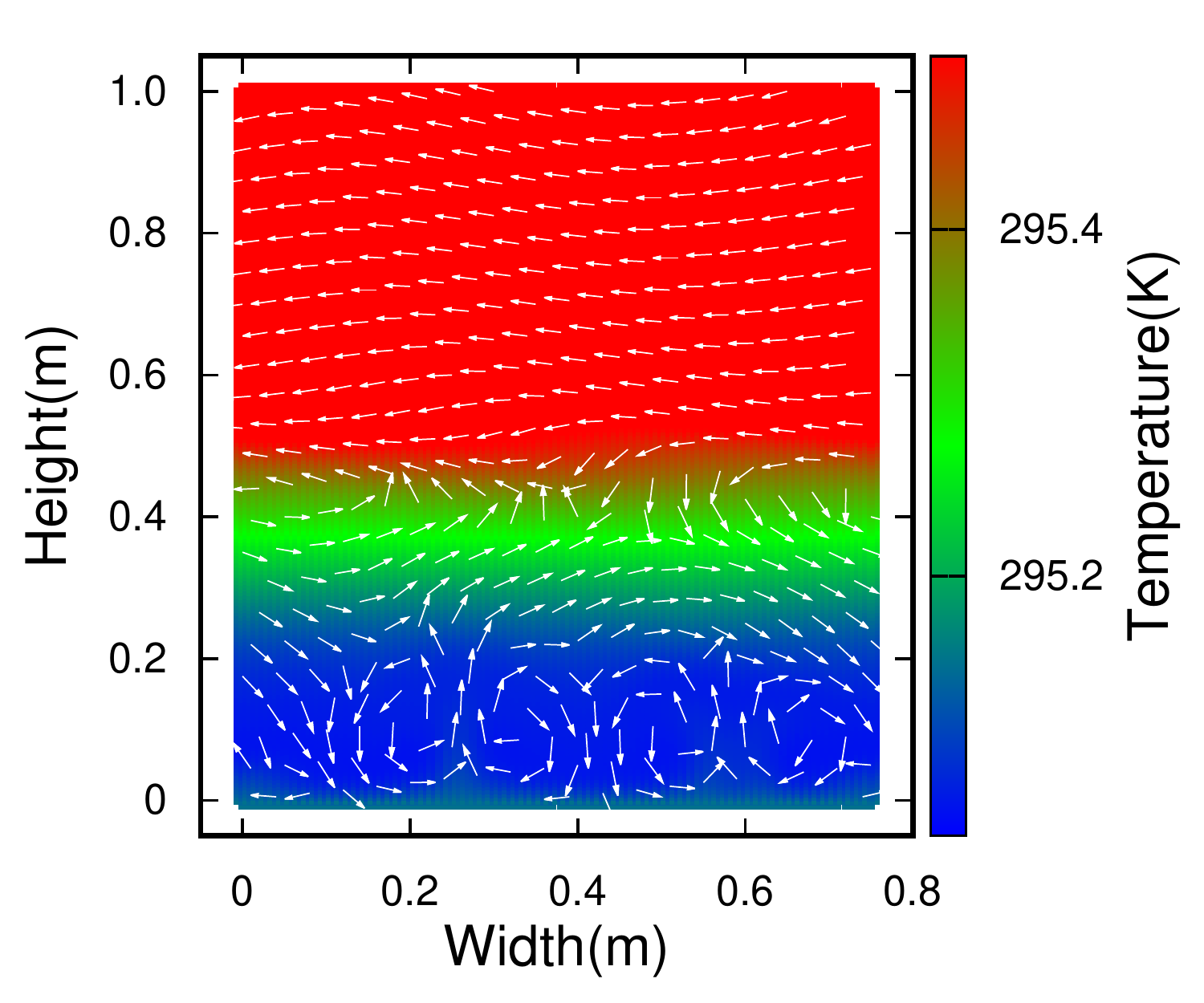} \\
		\small Case a) $\hspace{0.3cm} N_p = 0.1$
	\end{tabular} \qquad
	\begin{tabular}[b]{c}
		\includegraphics[width=.28\linewidth]{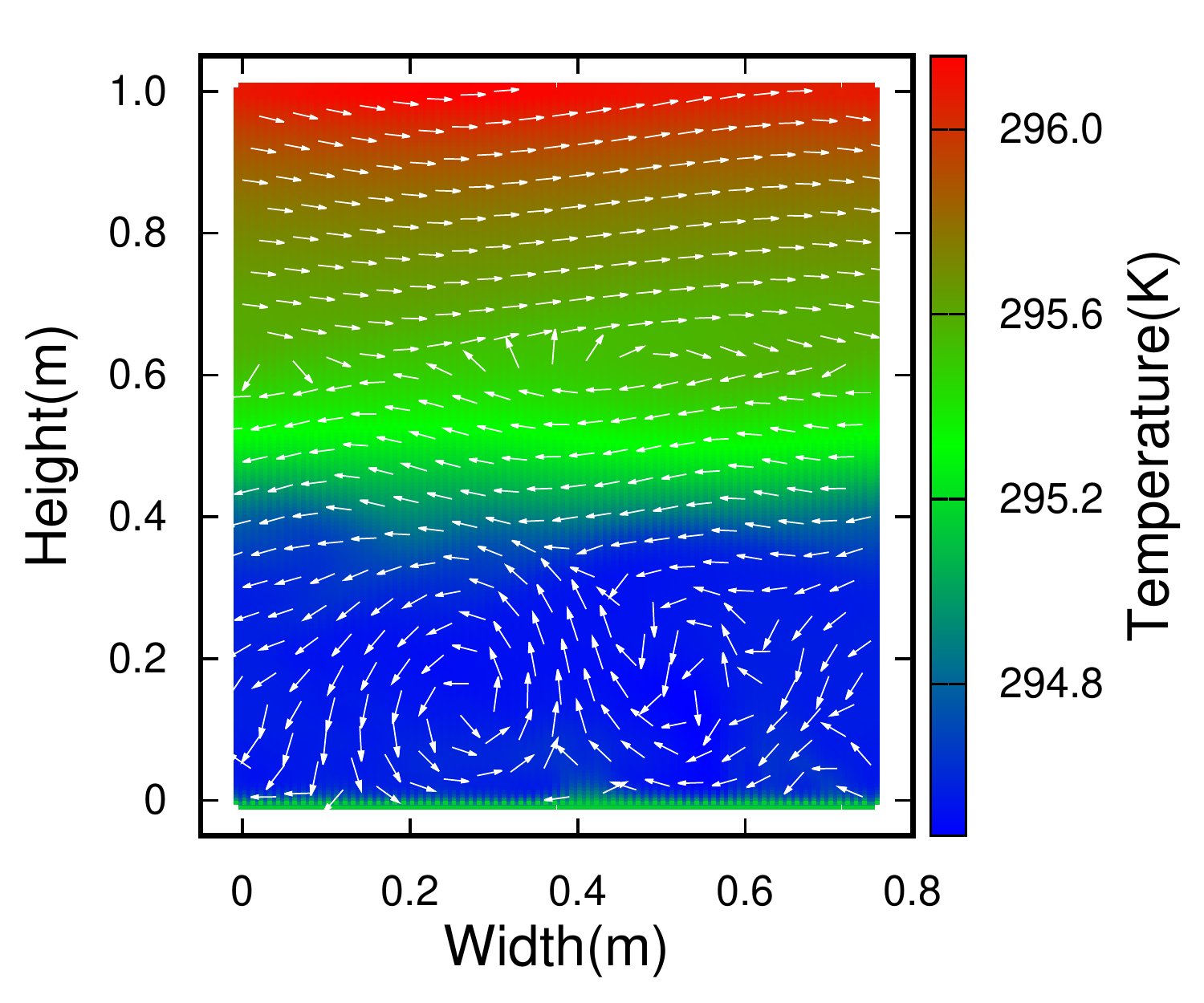} \\
		\small Case b) $\hspace{0.3cm} N_p = 1$
	\end{tabular} \qquad
	\begin{tabular}[b]{c}
		\includegraphics[width=.28\linewidth]{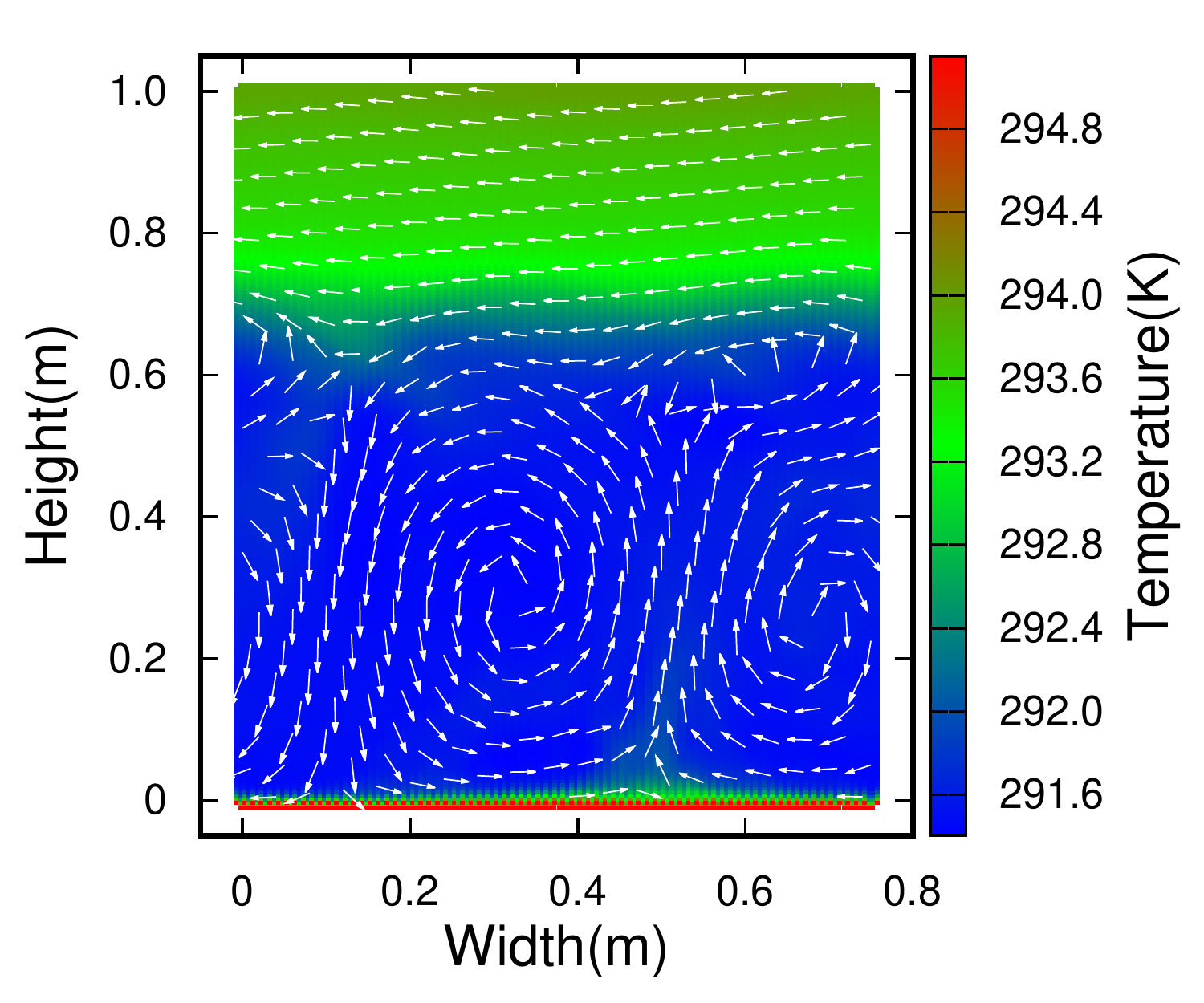} \\
		\small Case c) $\hspace{0.3cm} N_p = 10$
	\end{tabular}
	\caption{\label{density}Mean surface temperature profiles (at $t=585s$) plotted along with average velocity vectors (at $t=585s$). Both temperature and velocity vector, is averaged over the selected time interval $(570-600)s$, for all computational points, and is assigned as the temperature and velocity at $t=585s$.}
\end{figure*}

The local temperature and density data from the simulation is first horizontally averaged and then a running average in time (for $30$ seconds)  is carried out in order to obtain a temperature and density field, that is solely a function of height and time, devoid of turbulent fluctuations. 
Fig.\ref{temp} clearly shows that as the simulation progresses, the stably stratified (linear) initial condition quickly forms an LTM system with the minimum temperature occurring a few tens of centimeters above the lower boundary. This is similar to the LTM profile that is observed at night times under clear and calm conditions. 
Furthermore, the turbulent mixing in the inversion layer forms an almost isothermal region capped by a stably stratified linear profile on top (typical penetrative convection system). The plots depict the temperature profile for three intermediate times  across which the isothermal layer cools and grows progressively. A more detailed analysis of the entrainment of the non turbulent fluid and growth of the mixed layer, is carried out in the next section. 
Another observation is that the amount of cooling is clearly a function of the number density of the aerosol particles, which in all three cases in Fig \ref{temp} is varied using $N_p$. Case c) shows the maximum amount of cooling with the mixed layer temperature dropping to $291.6$K after $585$ seconds. Qualitatively, it is easy to notice a faster growth of the mixed layer in Case c), as opposed to Case a), where the mixed layer growth comes to a halt. 
Similarly the mean density plots are shown in Fig.\ref{density} and for all observation purposes mirror the mean temperature plots. In the next section we use the mean density profiles in order to systematically locate the interface in all the simulations.

\begin{figure}
	\includegraphics[scale=0.45]{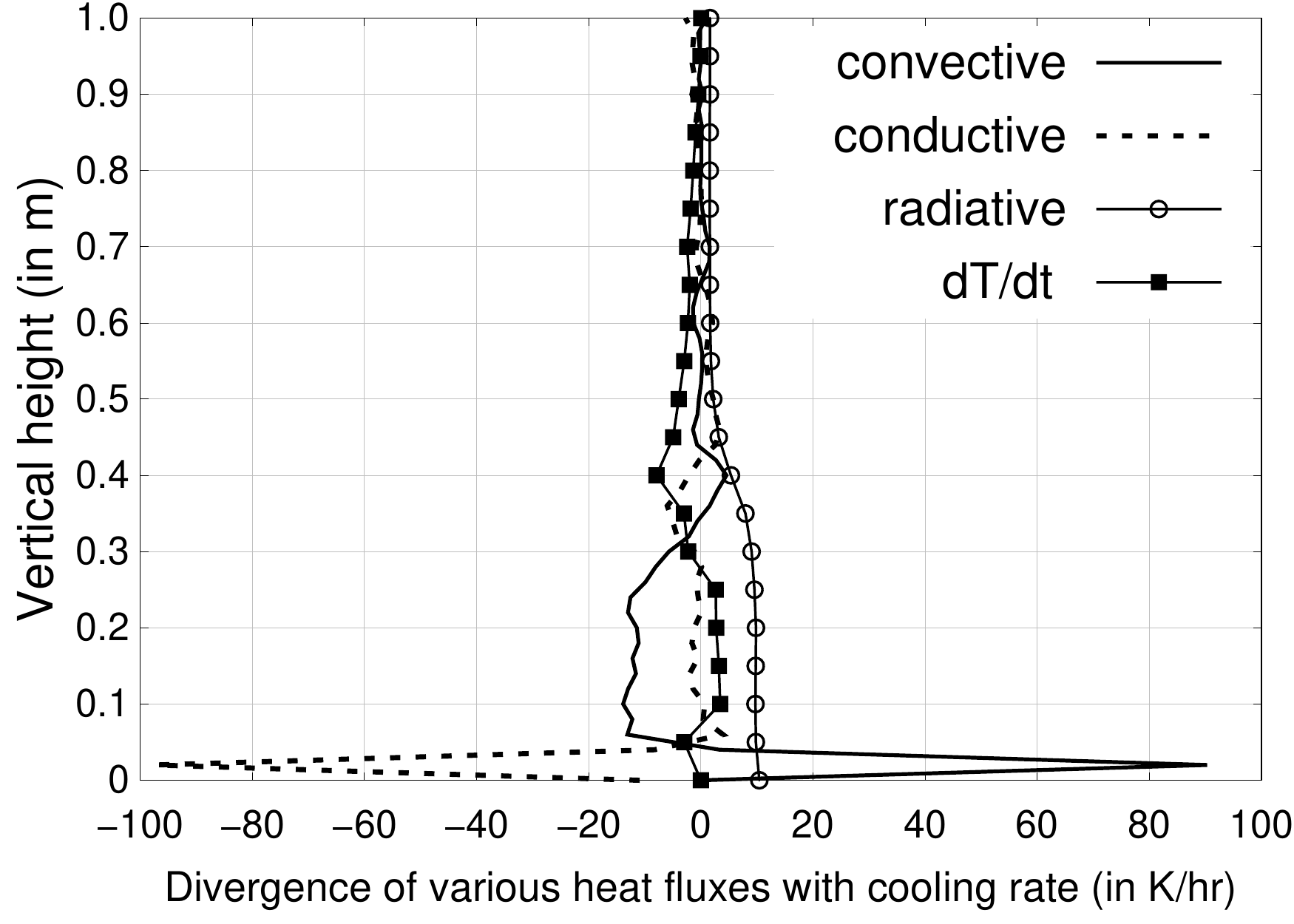}
	\caption{\label{heat_flux}Divergence of heat flux plotted as a function of height for, Case b), $N_p=1$. The respective terms of the heat equation (convection, conduction, radiation) are given in Eq.\ref{heat}. }
\end{figure}

The divergence of heat flux for conduction, convection and radiation is plotted as a function of height in Fig.\ref{heat_flux}, along with the cooling rate ($dT/dt$ in K/hr), as elements of the balanced heat equation, 
\begin{equation}
\frac{\partial T}{\partial t} + \underbrace{ \frac{1}{\rho C_p}\left( u\frac{\partial }{\partial x} + v \frac{\partial }{\partial y} \right)T }_{\text{convective}} = \underbrace{ \frac{k}{\rho C_p} \left( \frac{\partial^2}{\partial x^2} + \frac{\partial^2}{\partial y^2} \right) T}_{\text{conductive}} + \underbrace{\frac{q'''}{\rho C_p}}_{\text{radiative}}.
\label{heat}
\end{equation}
At a distance close to the ground, the sharp temperature gradient leads to a substantial conductive flux which is balanced majorly by the convective flux of the upward rising plumes. As we go higher, the conductive flux contribution becomes negligible and the radiative flux terms balances out the convective flux of downward moving plumes. The convective flux contribution also becomes negligible as we reach and cross the interface. As seen in Fig.\ref{heat_flux}, the heat equation is well balanced at all heights and the respective contribution of different heat fluxes follows the expected dynamics of a penetrative convection system. 

\subsection{\label{res2}Convectively driven mixed layer }

\begin{figure*}
	\centering
	\includegraphics[width=.6\linewidth]{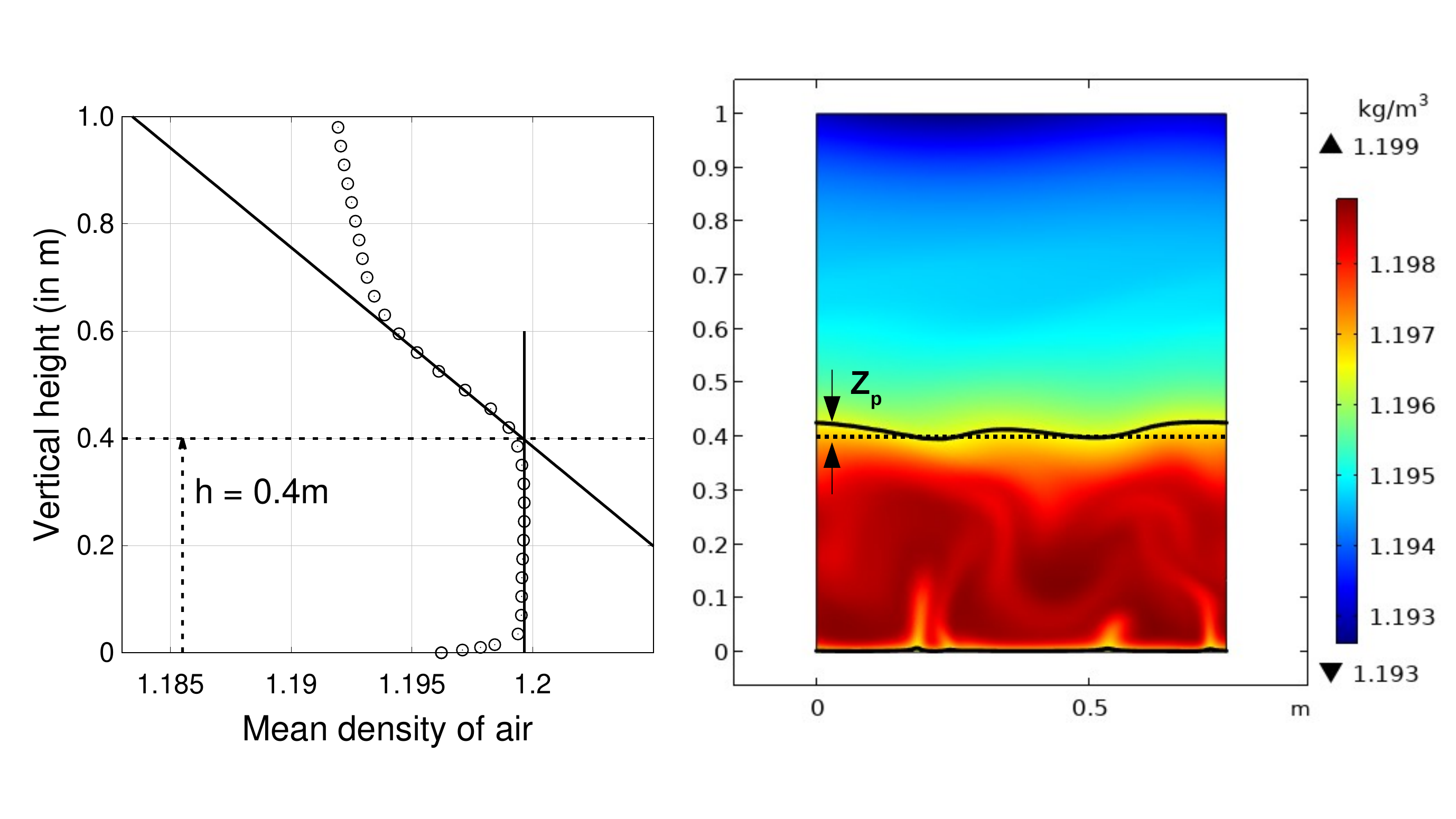} \\
	\caption{\label{find_interface1} Locating the interface using the density plots for the same instance. a) Horizontally averaged density data. b) Surface plot for density.}
\end{figure*}

\begin{figure*}
	\centering
	\includegraphics[width=.6\linewidth]{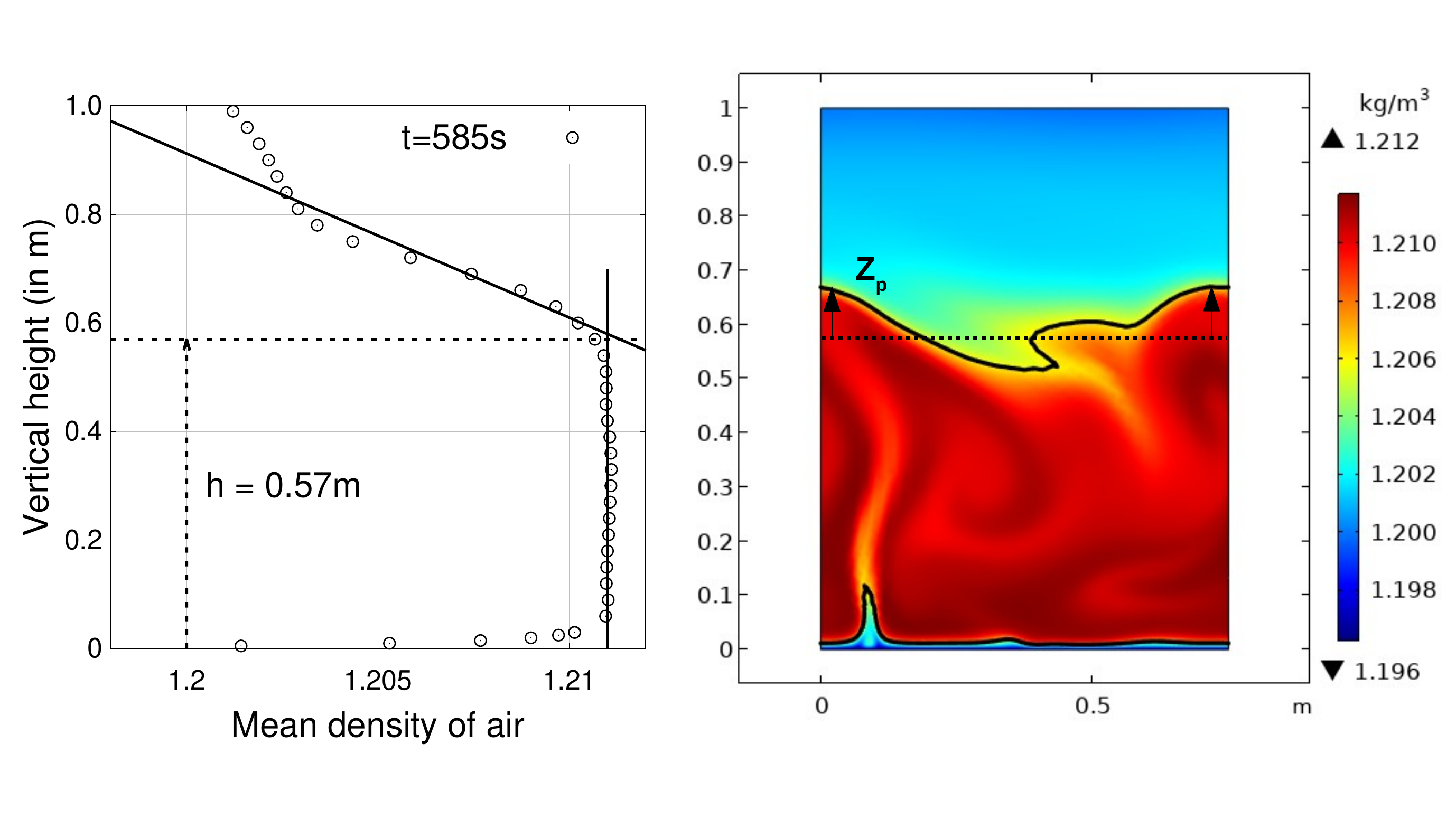} \\
	\caption{\label{find_interface2} Locating the interface using the density plots for the same instance. a) Horizontally averaged density data. b) Surface plot for density.}
\end{figure*}

\begin{figure}
	\includegraphics[scale=0.4]{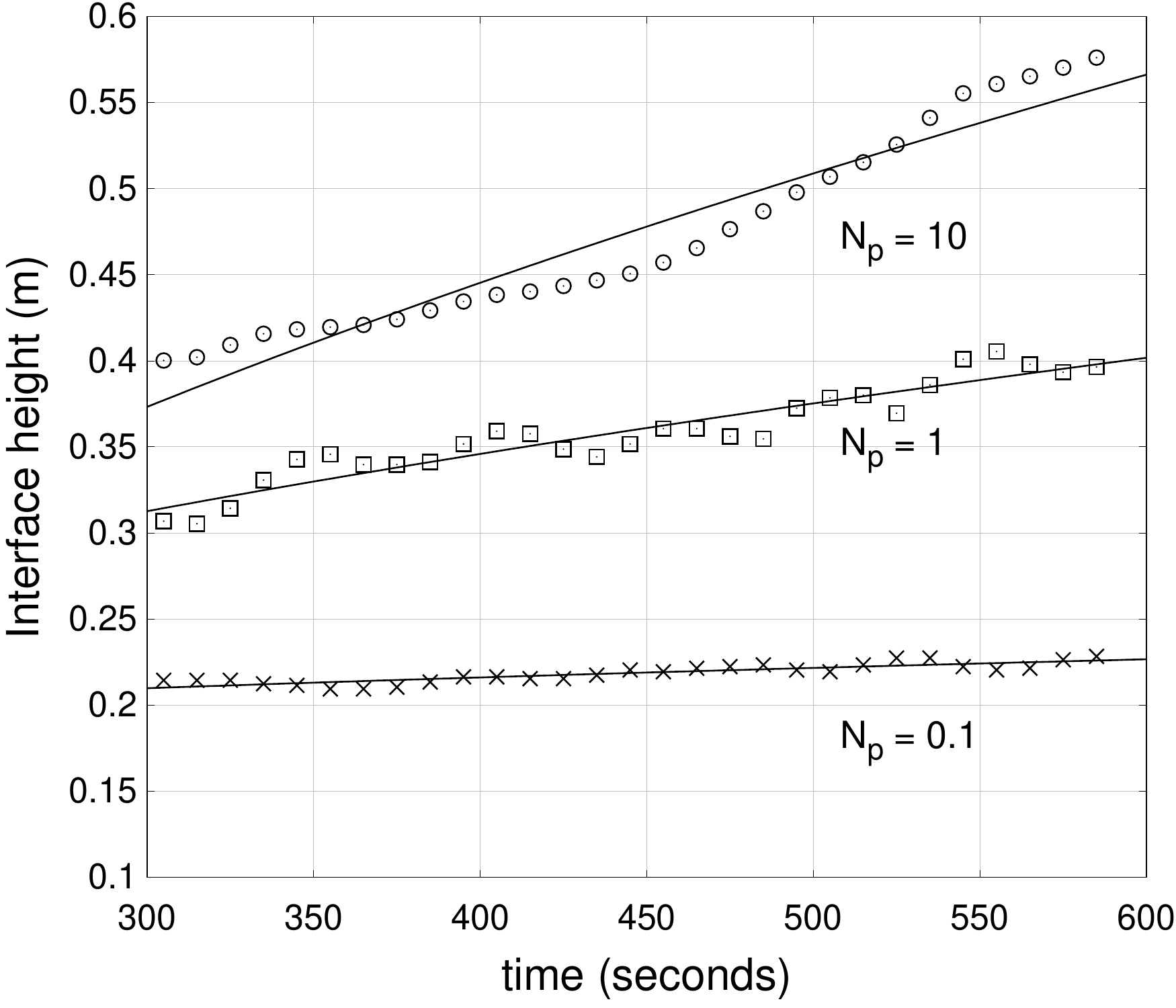}
	\caption{\label{hvst} Growth of the interface height as a function of time}
\end{figure}

Determining the mixed layer height is the first and foremost step for the analysis that follows in subsequent sections. By definition, the mixed layer is a region where turbulent mixing ensures a uniform density. Using this definition and the presence of a stably stratified (linear density) layer on top of the uniform density region, the mixed layer height($h$) is marked as seen in Fig.\ref{find_interface1} and Fig.\ref{find_interface2}.  
Once a mixed layer is formed, the mixed layer height is calculated at every time step  and plotted as a function of time in Fig.\ref{hvst}. The computational data points seem to very well fit  the solid lines that represent a $\propto \sqrt{t}$ dependence, as pointed out in previous studies \cite{kato}. The rate of change of mixed layer height is termed the entrainment velocity as is simply given by 
\begin{equation}
U_e = \frac{dh}{dt}
\end{equation}   

Once the mixed layer height is determined,the calculation of convective velocity scale ($U^*$) is quite straightforward.
\begin{equation}
U^* = \left[ \frac{g\beta Q_b h}{\rho C_p} \right]^{1/3}
\label{ustar}
\end{equation}
where, $g$ is the gravitational constant, $\rho$ is the density and $C_p$ is the specific heat for air at constant pressure. $Q_b$ is the bottom heat flux and is given by
\begin{equation}
Q_b = k \bar{\frac{dT}{dy}}\Big|_{y=0}
\end{equation}
where $k$ is the thermal conductivity of air.
Deardorff's \cite{deardorff1980} turbulent entrainment model for single diffusive systems (alongwith a number of follow up studies) proposes that the entrainment velocity ($U_e$) is a function of Richardson number $(Ri)$ and a convective velocity scale $(U^*)$ given by the relation, 
\begin{equation}
 U_e = C_1 U^* Ri^{-n},
 \label{main_eq}
\end{equation}
where, $C_1$ and $n$ are empirical constants. 

Richardson number is a widely used non-dimensional parameter that is used to predict the occurrence of fluid turbulence. It finds practical importance in weather forecasting and in investigating density and turbidity currents in oceans, lakes, and reservoirs. The definition of $Ri$ in the context of this study is,
\begin{equation}
Ri = \frac{g \left(\Delta \rho \right) h}{\rho (U^*)^2} 
\end{equation}
$\Delta \rho$ is the density jump across the interface. For the calculation of $Ri$, we use the  modified turbulent entrainment model \cite{sreenivas} to avoid running into a zero $Ri$ problem near the equilibrium, where the model predicts a physically unrealistic
entrainment velocity. The modified model divides $\Delta \rho$ into two components $\Delta \rho_1$ and $\Delta \rho_2$. $\Delta \rho_1$, is the density jump across the interface, and typical density profiles shown in Fig. \ref{find_interface1} and Fig. \ref{find_interface2}, suggest that $\Delta \rho_1 = 0$ for all the cases in this study.
Jump in density due to the presence of a density gradient is incorporated in the calculation of $\Delta \rho_2$.
$\Delta \rho_2$ is obtained by taking the product of the density gradient $\frac{d \rho}{dy}$ and a relevant length scale. The choice of relevant length scale is crucial to the problem. Choosing the mixed layer height$(h)$ for the length scale, as suggested by some \cite{zangrando}, leads to an overestimation of the resistance offered by the density gradient to the turbulent entrainment. 
Previous studies on penetrative convection \cite{deardorff1980,zeman, sreenivas} suggest using the penetration depth $(Z_p)$ as the relevant length scale. $Z_p$ is essentially the distance from the interface into the gradient zone over which the turbulent eddies have an
effect. 

\subsection{\label{res3} Entrainment Zone}

The outermost portion of the mixed layer where stably stratified fluid is entraining but
is not yet incorporated into the well-mixed layer is called the ‘entrainment zone ’.
The simplest assumption for $Z_p$, which has been used by Betts \cite{betts} is 
\begin{equation}
Z_p = \alpha h
\end{equation}
where $\alpha$ is a constant of order $0.3$. This approach doesn't seem to invoke any physical process to extract $Z_p$.

\begin{figure}
	\includegraphics[scale=0.4]{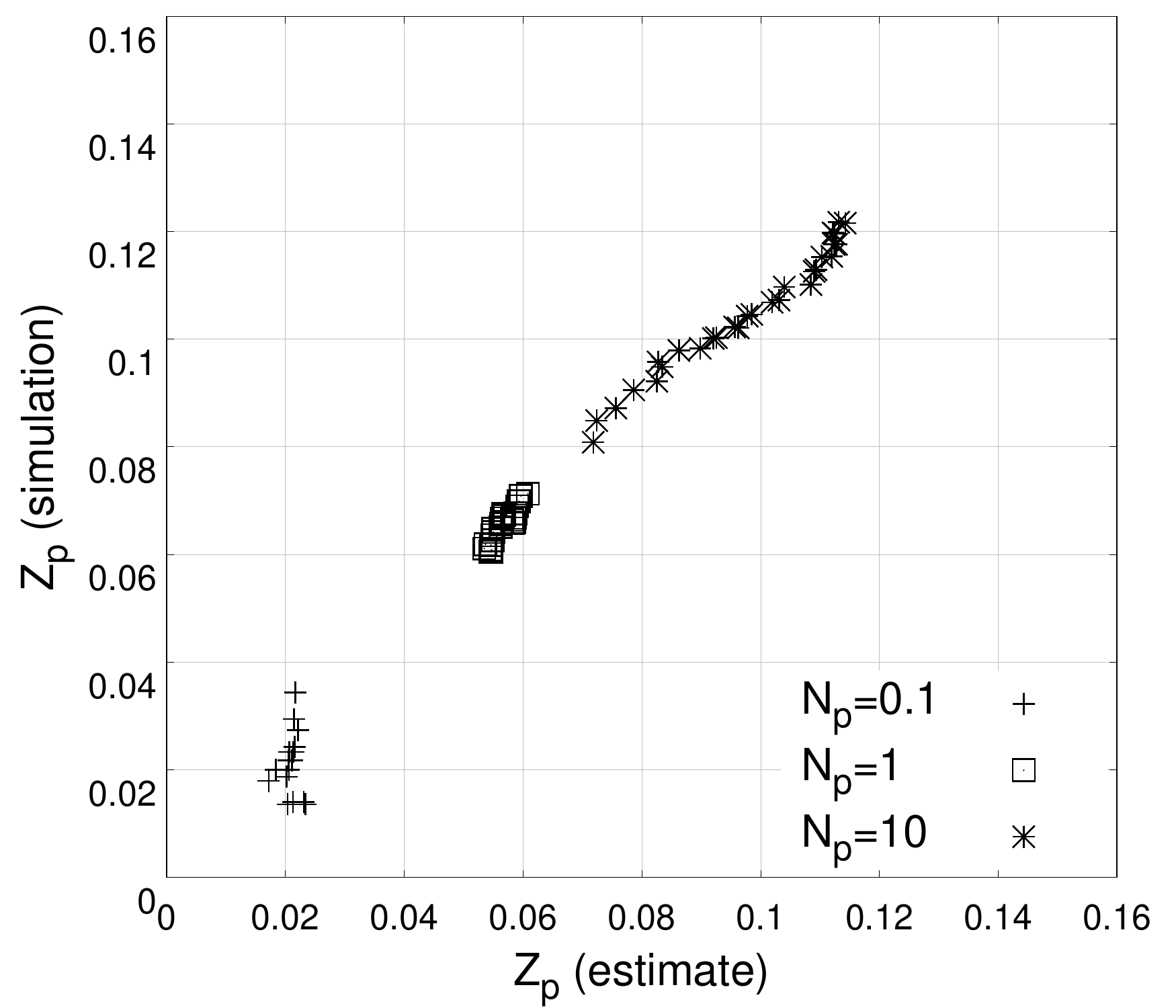}
	\caption{\label{zpvst} Penetration height ($Z_p$) from the simulation (using the protocol in Fig. \ref{find_interface1} and Fig. \ref{find_interface2}) is plotted against the estimated penetration depth as per Eq. \ref{zp}, for all the three test cases.}
\end{figure}

\begin{figure}
	\includegraphics[scale=0.4]{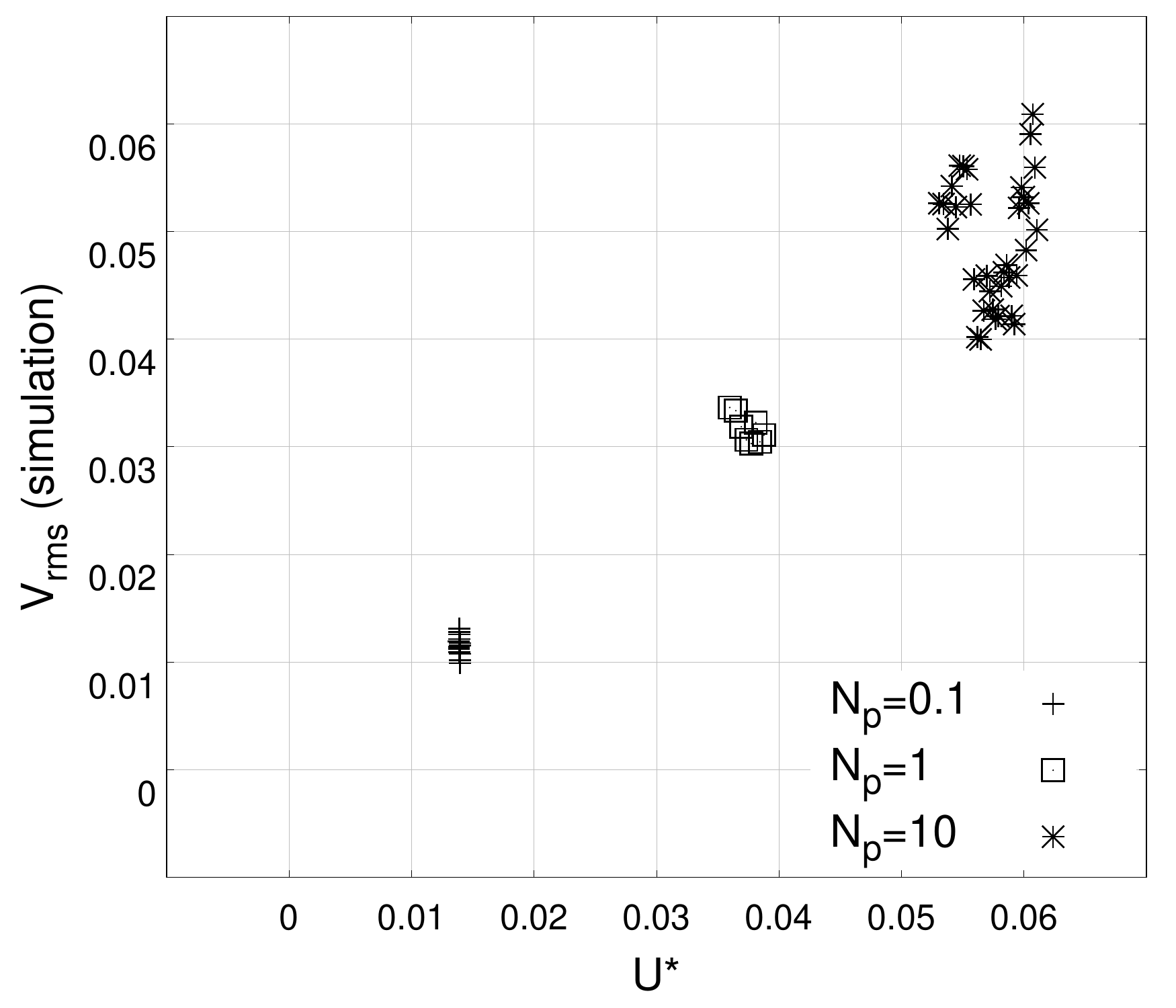}
	\caption{\label{vrms} Root mean squared vertical velocity ($V_{rms}$) from the simulation (calculated at the midpoint of the mixed layer) is plotted against the estimated convective velocity scale as per Eq. \ref{ustar}, for all the three test cases.}
\end{figure}

\begin{figure}
	\includegraphics[scale=0.4]{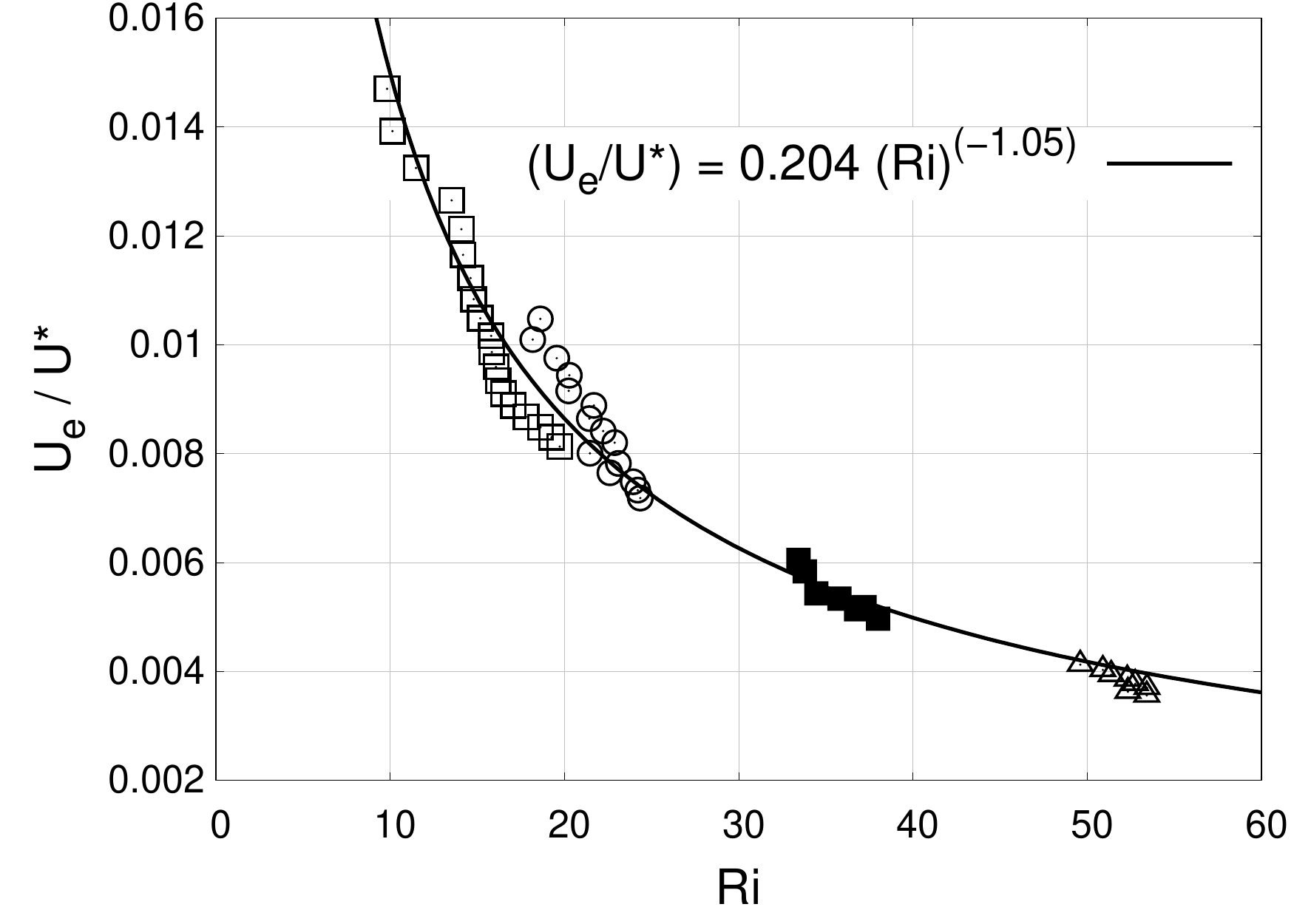}
	\caption{\label{Candn} $U_e/U^*$ plotted as a function of Richardson number. The data is extracted from the three sets of computational experiments and the the fit line gives $C_1 = 0.204$ and $n=1.05$}
\end{figure}

A balance of the kinetic energy per unit volume of the fluid $(KE = \rho(U^*)^2/2)$, in the mixed layer and potential energy gained in lifting a unit volume of fluid $(PE = \Delta \rho Z_p g)$ from the interface
into the gradient zone to a height of $Z_p$ gives,
\begin{equation}
Z_p = \sqrt{  \frac{\rho \left(U^*\right)^2}{2g \left(d\rho/dy \right) \big |_{y=h}}   }
\label{zp}
\end{equation}
The Richardson number for the problem is thus given by
\begin{equation}
Ri = \frac{g \left( \frac{d\rho}{dy}\Big|_{y=h} Z_p \right) h}{\rho (U^*)^2}
\end{equation}
Once the Richardson number is determined, we use Eq.\ref{main_eq} to find the empirical constants $C_1$ and $n$ as shown in Fig.\ref{Candn}. 
Now from our simulation, we check the validity of Eq.\ref{ustar} and \ref{zp} .
In the Fig.\ref{zpvst} we present the values of Zp extracted from present simulations
and compared that with the expression in Eq.\ref{zp}. Similarly
convective velocity scale ($V_{rms}$) observed in the simulations are
plotted against that estimated by Eq.\ref{ustar}, in Fig.\ref{vrms}.
Both these estimates compare well over $100$ times increase in aerosol
loading simulated here.

\section{Conclusions}
In the present work we have simulated LTM type of profiles
caused by the radiative cooling of aerosol laden nocturnal surface
layer. This cooling develops into a classical penetrative-convection
system. We have shown that in this situation, estimation of convective velocity
scale ($U^*$), entrainment zone thickness ($Z_p$) and entrainment
velocity (rate of growth of surface mixed layer) can be captured using
Deardorff model. However, one must use appropriate scale for $Z_p$
as given in Eq.\ref{zp}. Here we show that for $Z_p$, one should use height of
penetration of buoyant convective parcel into the stratified medium capping the
convective mixed layer. This model will be useful in predicting the growth
and decay of fog layer in the atmosphere.

\begin{acknowledgments}
We would like to thank Jawaharlal Nehru Centre for
Advanced Scientific Research, Bangalore, India, for the support.
\end{acknowledgments}

\section*{Data Availability Statement}
The data that support the findings of this study is openly available on gitlab. The link to the same is as follows, $\text{https://gitlab.com/shauryajncasr/penetrative-convection-in-nbl}$

\nocite{*}
\bibliography{references}

\providecommand{\noopsort}[1]{}\providecommand{\singleletter}[1]{#1}%
\begin{thebibliography}{33}%
\makeatletter
\providecommand \@ifxundefined [1]{%
 \@ifx{#1\undefined}
}%
\providecommand \@ifnum [1]{%
 \ifnum #1\expandafter \@firstoftwo
 \else \expandafter \@secondoftwo
 \fi
}%
\providecommand \@ifx [1]{%
 \ifx #1\expandafter \@firstoftwo
 \else \expandafter \@secondoftwo
 \fi
}%
\providecommand \natexlab [1]{#1}%
\providecommand \enquote  [1]{``#1''}%
\providecommand \bibnamefont  [1]{#1}%
\providecommand \bibfnamefont [1]{#1}%
\providecommand \citenamefont [1]{#1}%
\providecommand \href@noop [0]{\@secondoftwo}%
\providecommand \href [0]{\begingroup \@sanitize@url \@href}%
\providecommand \@href[1]{\@@startlink{#1}\@@href}%
\providecommand \@@href[1]{\endgroup#1\@@endlink}%
\providecommand \@sanitize@url [0]{\catcode `\\12\catcode `\$12\catcode
  `\&12\catcode `\#12\catcode `\^12\catcode `\_12\catcode `\%12\relax}%
\providecommand \@@startlink[1]{}%
\providecommand \@@endlink[0]{}%
\providecommand \url  [0]{\begingroup\@sanitize@url \@url }%
\providecommand \@url [1]{\endgroup\@href {#1}{\urlprefix }}%
\providecommand \urlprefix  [0]{URL }%
\providecommand \Eprint [0]{\href }%
\providecommand \doibase [0]{http://dx.doi.org/}%
\providecommand \selectlanguage [0]{\@gobble}%
\providecommand \bibinfo  [0]{\@secondoftwo}%
\providecommand \bibfield  [0]{\@secondoftwo}%
\providecommand \translation [1]{[#1]}%
\providecommand \BibitemOpen [0]{}%
\providecommand \bibitemStop [0]{}%
\providecommand \bibitemNoStop [0]{.\EOS\space}%
\providecommand \EOS [0]{\spacefactor3000\relax}%
\providecommand \BibitemShut  [1]{\csname bibitem#1\endcsname}%
\let\auto@bib@innerbib\@empty
\bibitem [{\citenamefont {Mukund}\ \emph {et~al.}(2014)\citenamefont {Mukund},
  \citenamefont {Singh}, \citenamefont {Ponnulakshmi}, \citenamefont
  {Subramanian},\ and\ \citenamefont {Sreenivas}}]{mukund}%
  \BibitemOpen
  \bibfield  {author} {\bibinfo {author} {\bibfnamefont {V.}~\bibnamefont
  {Mukund}}, \bibinfo {author} {\bibfnamefont {D.}~\bibnamefont {Singh}},
  \bibinfo {author} {\bibfnamefont {V.}~\bibnamefont {Ponnulakshmi}}, \bibinfo
  {author} {\bibfnamefont {G.}~\bibnamefont {Subramanian}}, \ and\ \bibinfo
  {author} {\bibfnamefont {K.}~\bibnamefont {Sreenivas}},\ }\bibfield  {title}
  {\enquote {\bibinfo {title} {Field and laboratory experiments on
  aerosol-induced cooling in the nocturnal boundary layer},}\ }\href@noop {}
  {\bibfield  {journal} {\bibinfo  {journal} {Quarterly Journal of the Royal
  Meteorological Society}\ }\textbf {\bibinfo {volume} {140}},\ \bibinfo
  {pages} {151--169} (\bibinfo {year} {2014})}\BibitemShut {NoStop}%
\bibitem [{\citenamefont {Singh}\ \emph {et~al.}(2013)\citenamefont {Singh},
  \citenamefont {Ponnulakshami}, \citenamefont {Mukund}, \citenamefont
  {Subramanian},\ and\ \citenamefont {Sreenivas}}]{singh}%
  \BibitemOpen
  \bibfield  {author} {\bibinfo {author} {\bibfnamefont {D.}~\bibnamefont
  {Singh}}, \bibinfo {author} {\bibfnamefont {V.}~\bibnamefont
  {Ponnulakshami}}, \bibinfo {author} {\bibfnamefont {V.}~\bibnamefont
  {Mukund}}, \bibinfo {author} {\bibfnamefont {G.}~\bibnamefont {Subramanian}},
  \ and\ \bibinfo {author} {\bibfnamefont {K.}~\bibnamefont {Sreenivas}},\
  }\bibfield  {title} {\enquote {\bibinfo {title} {Radiation forcing by the
  atmospheric aerosols in the nocturnal boundary layer},}\ }in\ \href@noop {}
  {\emph {\bibinfo {booktitle} {AIP Conference Proceedings}}},\ Vol.\ \bibinfo
  {volume} {1531}\ (\bibinfo {organization} {American Institute of Physics},\
  \bibinfo {year} {2013})\ pp.\ \bibinfo {pages} {596--599}\BibitemShut
  {NoStop}%
\bibitem [{\citenamefont {Ramdas}\ and\ \citenamefont
  {Atmanathan}(1932)}]{ramdas}%
  \BibitemOpen
  \bibfield  {author} {\bibinfo {author} {\bibfnamefont {L.}~\bibnamefont
  {Ramdas}}\ and\ \bibinfo {author} {\bibfnamefont {S.}~\bibnamefont
  {Atmanathan}},\ }\bibfield  {title} {\enquote {\bibinfo {title} {The vertical
  distribution of air temperature near the ground at night},}\ }\href@noop {}
  {\bibfield  {journal} {\bibinfo  {journal} {Beit. Geophys}\ }\textbf
  {\bibinfo {volume} {37}},\ \bibinfo {pages} {116--117} (\bibinfo {year}
  {1932})}\BibitemShut {NoStop}%
\bibitem [{\citenamefont {Blay-Carreras}\ \emph {et~al.}(2015)\citenamefont
  {Blay-Carreras}, \citenamefont {Pardyjak}, \citenamefont {Pino},
  \citenamefont {Hoch}, \citenamefont {Cuxart}, \citenamefont {Mart{\'\i}nez},\
  and\ \citenamefont {Reuder}}]{blay2015}%
  \BibitemOpen
  \bibfield  {author} {\bibinfo {author} {\bibfnamefont {E.}~\bibnamefont
  {Blay-Carreras}}, \bibinfo {author} {\bibfnamefont {E.}~\bibnamefont
  {Pardyjak}}, \bibinfo {author} {\bibfnamefont {D.}~\bibnamefont {Pino}},
  \bibinfo {author} {\bibfnamefont {S.}~\bibnamefont {Hoch}}, \bibinfo {author}
  {\bibfnamefont {J.}~\bibnamefont {Cuxart}}, \bibinfo {author} {\bibfnamefont
  {D.}~\bibnamefont {Mart{\'\i}nez}}, \ and\ \bibinfo {author} {\bibfnamefont
  {J.}~\bibnamefont {Reuder}},\ }\bibfield  {title} {\enquote {\bibinfo {title}
  {Lifted temperature minimum during the atmospheric evening transition},}\
  }\href@noop {} {\bibfield  {journal} {\bibinfo  {journal} {Atmospheric
  Chemistry and Physics}\ }\textbf {\bibinfo {volume} {15}},\ \bibinfo {pages}
  {6981--6991} (\bibinfo {year} {2015})}\BibitemShut {NoStop}%
\bibitem [{\citenamefont {Townsend}(1964)}]{townsend}%
  \BibitemOpen
  \bibfield  {author} {\bibinfo {author} {\bibfnamefont {A.}~\bibnamefont
  {Townsend}},\ }\bibfield  {title} {\enquote {\bibinfo {title} {Natural
  convection in water over an ice surface},}\ }\href@noop {} {\bibfield
  {journal} {\bibinfo  {journal} {Quarterly Journal of the Royal Meteorological
  Society}\ }\textbf {\bibinfo {volume} {90}},\ \bibinfo {pages} {248--259}
  (\bibinfo {year} {1964})}\BibitemShut {NoStop}%
\bibitem [{\citenamefont {Adrian}(1975)}]{adrian}%
  \BibitemOpen
  \bibfield  {author} {\bibinfo {author} {\bibfnamefont {R.}~\bibnamefont
  {Adrian}},\ }\bibfield  {title} {\enquote {\bibinfo {title} {Turbulent
  convection in water over ice},}\ }\href@noop {} {\bibfield  {journal}
  {\bibinfo  {journal} {Journal of Fluid Mechanics}\ }\textbf {\bibinfo
  {volume} {69}},\ \bibinfo {pages} {753--781} (\bibinfo {year}
  {1975})}\BibitemShut {NoStop}%
\bibitem [{\citenamefont {Veronis}(1963)}]{veronis}%
  \BibitemOpen
  \bibfield  {author} {\bibinfo {author} {\bibfnamefont {G.}~\bibnamefont
  {Veronis}},\ }\bibfield  {title} {\enquote {\bibinfo {title} {Penetrative
  convection.}}\ }\href@noop {} {\bibfield  {journal} {\bibinfo  {journal} {The
  Astrophysical Journal}\ }\textbf {\bibinfo {volume} {137}},\ \bibinfo {pages}
  {641} (\bibinfo {year} {1963})}\BibitemShut {NoStop}%
\bibitem [{\citenamefont {Driedonks}(1982)}]{driedonks}%
  \BibitemOpen
  \bibfield  {author} {\bibinfo {author} {\bibfnamefont {A.}~\bibnamefont
  {Driedonks}},\ }\bibfield  {title} {\enquote {\bibinfo {title} {Models and
  observations of the growth of the atmospheric boundary layer},}\ }\href@noop
  {} {\bibfield  {journal} {\bibinfo  {journal} {Boundary-Layer Meteorology}\
  }\textbf {\bibinfo {volume} {23}},\ \bibinfo {pages} {283--306} (\bibinfo
  {year} {1982})}\BibitemShut {NoStop}%
\bibitem [{\citenamefont {Leighton}(1963)}]{leighton}%
  \BibitemOpen
  \bibfield  {author} {\bibinfo {author} {\bibfnamefont {R.~B.}\ \bibnamefont
  {Leighton}},\ }\bibfield  {title} {\enquote {\bibinfo {title} {The solar
  granulation},}\ }\href@noop {} {\bibfield  {journal} {\bibinfo  {journal}
  {Annual review of astronomy and astrophysics}\ }\textbf {\bibinfo {volume}
  {1}},\ \bibinfo {pages} {19--40} (\bibinfo {year} {1963})}\BibitemShut
  {NoStop}%
\bibitem [{\citenamefont {Simpson}\ \emph {et~al.}(1965)\citenamefont
  {Simpson}, \citenamefont {Simpson}, \citenamefont {Andrews},\ and\
  \citenamefont {Eaton}}]{simpson}%
  \BibitemOpen
  \bibfield  {author} {\bibinfo {author} {\bibfnamefont {J.}~\bibnamefont
  {Simpson}}, \bibinfo {author} {\bibfnamefont {R.~H.}\ \bibnamefont
  {Simpson}}, \bibinfo {author} {\bibfnamefont {D.~A.}\ \bibnamefont
  {Andrews}}, \ and\ \bibinfo {author} {\bibfnamefont {M.~A.}\ \bibnamefont
  {Eaton}},\ }\bibfield  {title} {\enquote {\bibinfo {title} {Experimental
  cumulus dynamics},}\ }\href@noop {} {\bibfield  {journal} {\bibinfo
  {journal} {Reviews of Geophysics}\ }\textbf {\bibinfo {volume} {3}},\
  \bibinfo {pages} {387--431} (\bibinfo {year} {1965})}\BibitemShut {NoStop}%
\bibitem [{\citenamefont {Deardorff}, \citenamefont {Willis},\ and\
  \citenamefont {Stockton}(1980)}]{deardorff1980}%
  \BibitemOpen
  \bibfield  {author} {\bibinfo {author} {\bibfnamefont {J.}~\bibnamefont
  {Deardorff}}, \bibinfo {author} {\bibfnamefont {G.}~\bibnamefont {Willis}}, \
  and\ \bibinfo {author} {\bibfnamefont {B.}~\bibnamefont {Stockton}},\
  }\bibfield  {title} {\enquote {\bibinfo {title} {Laboratory studies of the
  entrainment zone of a convectively mixed layer},}\ }\href@noop {} {\bibfield
  {journal} {\bibinfo  {journal} {Journal of Fluid Mechanics}\ }\textbf
  {\bibinfo {volume} {100}},\ \bibinfo {pages} {41--64} (\bibinfo {year}
  {1980})}\BibitemShut {NoStop}%
\bibitem [{\citenamefont {Willis}\ and\ \citenamefont
  {Deardorff}(1974)}]{willis1974}%
  \BibitemOpen
  \bibfield  {author} {\bibinfo {author} {\bibfnamefont {G.}~\bibnamefont
  {Willis}}\ and\ \bibinfo {author} {\bibfnamefont {J.}~\bibnamefont
  {Deardorff}},\ }\bibfield  {title} {\enquote {\bibinfo {title} {A laboratory
  model of the unstable planetary boundary layer},}\ }\href@noop {} {\bibfield
  {journal} {\bibinfo  {journal} {Journal of Atmospheric Sciences}\ }\textbf
  {\bibinfo {volume} {31}},\ \bibinfo {pages} {1297--1307} (\bibinfo {year}
  {1974})}\BibitemShut {NoStop}%
\bibitem [{\citenamefont {Kumar}(1989)}]{kumar}%
  \BibitemOpen
  \bibfield  {author} {\bibinfo {author} {\bibfnamefont {R.}~\bibnamefont
  {Kumar}},\ }\bibfield  {title} {\enquote {\bibinfo {title} {Laboratory
  studies of thermal convection in the interface under a stable layer},}\
  }\href@noop {} {\bibfield  {journal} {\bibinfo  {journal} {International
  journal of heat and mass transfer}\ }\textbf {\bibinfo {volume} {32}},\
  \bibinfo {pages} {735--749} (\bibinfo {year} {1989})}\BibitemShut {NoStop}%
\bibitem [{\citenamefont {Sreenivas}, \citenamefont {Arakeri},\ and\
  \citenamefont {Srinivasan}(1995)}]{sreenivas}%
  \BibitemOpen
  \bibfield  {author} {\bibinfo {author} {\bibfnamefont {K.}~\bibnamefont
  {Sreenivas}}, \bibinfo {author} {\bibfnamefont {J.~H.}\ \bibnamefont
  {Arakeri}}, \ and\ \bibinfo {author} {\bibfnamefont {J.}~\bibnamefont
  {Srinivasan}},\ }\bibfield  {title} {\enquote {\bibinfo {title} {Modeling the
  dynamics of the mixed layer in solar ponds},}\ }\href@noop {} {\bibfield
  {journal} {\bibinfo  {journal} {Solar energy}\ }\textbf {\bibinfo {volume}
  {54}},\ \bibinfo {pages} {193--202} (\bibinfo {year} {1995})}\BibitemShut
  {NoStop}%
\bibitem [{\citenamefont {Deardorff}, \citenamefont {Willis},\ and\
  \citenamefont {Lilly}(1969)}]{deardorff}%
  \BibitemOpen
  \bibfield  {author} {\bibinfo {author} {\bibfnamefont {J.~W.}\ \bibnamefont
  {Deardorff}}, \bibinfo {author} {\bibfnamefont {G.~E.}\ \bibnamefont
  {Willis}}, \ and\ \bibinfo {author} {\bibfnamefont {D.~K.}\ \bibnamefont
  {Lilly}},\ }\bibfield  {title} {\enquote {\bibinfo {title} {Laboratory
  investigation of non-steady penetrative convection},}\ }\href@noop {}
  {\bibfield  {journal} {\bibinfo  {journal} {Journal of Fluid Mechanics}\
  }\textbf {\bibinfo {volume} {35}},\ \bibinfo {pages} {7--31} (\bibinfo {year}
  {1969})}\BibitemShut {NoStop}%
\bibitem [{\citenamefont {Fernando}\ and\ \citenamefont
  {Little}(1990)}]{fernando}%
  \BibitemOpen
  \bibfield  {author} {\bibinfo {author} {\bibfnamefont {H.}~\bibnamefont
  {Fernando}}\ and\ \bibinfo {author} {\bibfnamefont {L.~J.}\ \bibnamefont
  {Little}},\ }\bibfield  {title} {\enquote {\bibinfo {title}
  {Molecular-diffusive effects in penetrative convection},}\ }\href@noop {}
  {\bibfield  {journal} {\bibinfo  {journal} {Physics of Fluids A: Fluid
  Dynamics}\ }\textbf {\bibinfo {volume} {2}},\ \bibinfo {pages} {1592--1596}
  (\bibinfo {year} {1990})}\BibitemShut {NoStop}%
\bibitem [{\citenamefont {Hutchison}\ and\ \citenamefont
  {Richards}(1999)}]{hutchison}%
  \BibitemOpen
  \bibfield  {author} {\bibinfo {author} {\bibfnamefont {J.}~\bibnamefont
  {Hutchison}}\ and\ \bibinfo {author} {\bibfnamefont {R.}~\bibnamefont
  {Richards}},\ }\bibfield  {title} {\enquote {\bibinfo {title} {Effect of
  nongray gas radiation on thermal stability in carbon dioxide},}\ }\href@noop
  {} {\bibfield  {journal} {\bibinfo  {journal} {Journal of thermophysics and
  heat transfer}\ }\textbf {\bibinfo {volume} {13}},\ \bibinfo {pages} {25--32}
  (\bibinfo {year} {1999})}\BibitemShut {NoStop}%
\bibitem [{\citenamefont {M{\"o}hlmann}\ \emph {et~al.}(2009)\citenamefont
  {M{\"o}hlmann}, \citenamefont {Niemand}, \citenamefont {Formisano},
  \citenamefont {Savij{\"a}rvi},\ and\ \citenamefont {Wolkenberg}}]{mars}%
  \BibitemOpen
  \bibfield  {author} {\bibinfo {author} {\bibfnamefont {D.~T.}\ \bibnamefont
  {M{\"o}hlmann}}, \bibinfo {author} {\bibfnamefont {M.}~\bibnamefont
  {Niemand}}, \bibinfo {author} {\bibfnamefont {V.}~\bibnamefont {Formisano}},
  \bibinfo {author} {\bibfnamefont {H.}~\bibnamefont {Savij{\"a}rvi}}, \ and\
  \bibinfo {author} {\bibfnamefont {P.}~\bibnamefont {Wolkenberg}},\ }\bibfield
   {title} {\enquote {\bibinfo {title} {Fog phenomena on mars},}\ }\href@noop
  {} {\bibfield  {journal} {\bibinfo  {journal} {Planetary and Space Science}\
  }\textbf {\bibinfo {volume} {57}},\ \bibinfo {pages} {1987--1992} (\bibinfo
  {year} {2009})}\BibitemShut {NoStop}%
\bibitem [{\citenamefont {Singh}(2013)}]{dhiraj_thesis}%
  \BibitemOpen
  \bibfield  {author} {\bibinfo {author} {\bibfnamefont {D.~K.}\ \bibnamefont
  {Singh}},\ }\bibfield  {title} {\enquote {\bibinfo {title} {The impact of
  aerosols and land surface properties on the lifted temperature minimum in the
  nocturnal atmospheric boundary layer - field and laboratory experiments},}\
  }\href@noop {} {\  (\bibinfo {year} {2013})}\BibitemShut {NoStop}%
\bibitem [{\citenamefont {Soulsby}(1997)}]{soulsby}%
  \BibitemOpen
  \bibfield  {author} {\bibinfo {author} {\bibfnamefont {R.}~\bibnamefont
  {Soulsby}},\ }\bibfield  {title} {\enquote {\bibinfo {title} {Dynamics of
  marine sands},}\ }\href@noop {} {\  (\bibinfo {year} {1997})}\BibitemShut
  {NoStop}%
\bibitem [{\citenamefont {Nielsen}\ and\ \citenamefont
  {Teakle}(2004)}]{nielsen}%
  \BibitemOpen
  \bibfield  {author} {\bibinfo {author} {\bibfnamefont {P.}~\bibnamefont
  {Nielsen}}\ and\ \bibinfo {author} {\bibfnamefont {I.~A.}\ \bibnamefont
  {Teakle}},\ }\bibfield  {title} {\enquote {\bibinfo {title} {Turbulent
  diffusion of momentum and suspended particles: A finite-mixing-length
  theory},}\ }\href@noop {} {\bibfield  {journal} {\bibinfo  {journal} {Physics
  of fluids}\ }\textbf {\bibinfo {volume} {16}},\ \bibinfo {pages} {2342--2348}
  (\bibinfo {year} {2004})}\BibitemShut {NoStop}%
\bibitem [{\citenamefont {Devara}\ and\ \citenamefont
  {Raj}(1993)}]{devara1993}%
  \BibitemOpen
  \bibfield  {author} {\bibinfo {author} {\bibfnamefont {P.}~\bibnamefont
  {Devara}}\ and\ \bibinfo {author} {\bibfnamefont {P.~E.}\ \bibnamefont
  {Raj}},\ }\bibfield  {title} {\enquote {\bibinfo {title} {Lidar measurements
  of aerosols in the tropical atmosphere},}\ }\href@noop {} {\bibfield
  {journal} {\bibinfo  {journal} {Advances in atmospheric sciences}\ }\textbf
  {\bibinfo {volume} {10}},\ \bibinfo {pages} {365--378} (\bibinfo {year}
  {1993})}\BibitemShut {NoStop}%
\bibitem [{\citenamefont {Raj}\ \emph {et~al.}(1997)\citenamefont {Raj},
  \citenamefont {Devara}, \citenamefont {Maheskumar}, \citenamefont
  {Pandithurai},\ and\ \citenamefont {Dani}}]{devara1997}%
  \BibitemOpen
  \bibfield  {author} {\bibinfo {author} {\bibfnamefont {P.~E.}\ \bibnamefont
  {Raj}}, \bibinfo {author} {\bibfnamefont {P.}~\bibnamefont {Devara}},
  \bibinfo {author} {\bibfnamefont {R.}~\bibnamefont {Maheskumar}}, \bibinfo
  {author} {\bibfnamefont {G.}~\bibnamefont {Pandithurai}}, \ and\ \bibinfo
  {author} {\bibfnamefont {K.}~\bibnamefont {Dani}},\ }\bibfield  {title}
  {\enquote {\bibinfo {title} {Lidar measurements of aerosol column content in
  an urban nocturnal boundary layer},}\ }\href@noop {} {\bibfield  {journal}
  {\bibinfo  {journal} {Atmospheric research}\ }\textbf {\bibinfo {volume}
  {45}},\ \bibinfo {pages} {201--216} (\bibinfo {year} {1997})}\BibitemShut
  {NoStop}%
\bibitem [{\citenamefont {Kato}\ and\ \citenamefont {Phillips}(1969)}]{kato}%
  \BibitemOpen
  \bibfield  {author} {\bibinfo {author} {\bibfnamefont {H.}~\bibnamefont
  {Kato}}\ and\ \bibinfo {author} {\bibfnamefont {O.}~\bibnamefont
  {Phillips}},\ }\bibfield  {title} {\enquote {\bibinfo {title} {On the
  penetration of a turbulent layer into stratified fluid},}\ }\href@noop {}
  {\bibfield  {journal} {\bibinfo  {journal} {Journal of Fluid Mechanics}\
  }\textbf {\bibinfo {volume} {37}},\ \bibinfo {pages} {643--655} (\bibinfo
  {year} {1969})}\BibitemShut {NoStop}%
\bibitem [{\citenamefont {Zangrando}\ and\ \citenamefont
  {Fernando}(1991)}]{zangrando}%
  \BibitemOpen
  \bibfield  {author} {\bibinfo {author} {\bibfnamefont {F.}~\bibnamefont
  {Zangrando}}\ and\ \bibinfo {author} {\bibfnamefont {H.}~\bibnamefont
  {Fernando}},\ }\bibfield  {title} {\enquote {\bibinfo {title} {A predictive
  model for the migration of double-diffusive interfaces},}\ }\href@noop {} {\
  (\bibinfo {year} {1991})}\BibitemShut {NoStop}%
\bibitem [{\citenamefont {Zeman}\ and\ \citenamefont {Tennekes}(1977)}]{zeman}%
  \BibitemOpen
  \bibfield  {author} {\bibinfo {author} {\bibfnamefont {O.}~\bibnamefont
  {Zeman}}\ and\ \bibinfo {author} {\bibfnamefont {H.}~\bibnamefont
  {Tennekes}},\ }\bibfield  {title} {\enquote {\bibinfo {title}
  {Parameterization of the turbulent energy budget at the top of the daytime
  atmospheric boundary layer},}\ }\href@noop {} {\bibfield  {journal} {\bibinfo
   {journal} {Journal of Atmospheric Sciences}\ }\textbf {\bibinfo {volume}
  {34}},\ \bibinfo {pages} {111--123} (\bibinfo {year} {1977})}\BibitemShut
  {NoStop}%
\bibitem [{\citenamefont {Betts}(1976)}]{betts}%
  \BibitemOpen
  \bibfield  {author} {\bibinfo {author} {\bibfnamefont {A.~K.}\ \bibnamefont
  {Betts}},\ }\bibfield  {title} {\enquote {\bibinfo {title} {The thermodynamic
  transformation of the tropical subcloud layer by precipitation and
  downdrafts},}\ }\href@noop {} {\bibfield  {journal} {\bibinfo  {journal}
  {Journal of atmospheric sciences}\ }\textbf {\bibinfo {volume} {33}},\
  \bibinfo {pages} {1008--1020} (\bibinfo {year} {1976})}\BibitemShut {NoStop}%
\bibitem [{\citenamefont {Musman}(1968)}]{musman}%
  \BibitemOpen
  \bibfield  {author} {\bibinfo {author} {\bibfnamefont {S.}~\bibnamefont
  {Musman}},\ }\bibfield  {title} {\enquote {\bibinfo {title} {Penetrative
  convection},}\ }\href@noop {} {\bibfield  {journal} {\bibinfo  {journal}
  {Journal of Fluid Mechanics}\ }\textbf {\bibinfo {volume} {31}},\ \bibinfo
  {pages} {343--360} (\bibinfo {year} {1968})}\BibitemShut {NoStop}%
\bibitem [{\citenamefont {Willis}\ and\ \citenamefont
  {Deardorff}(1979)}]{willis}%
  \BibitemOpen
  \bibfield  {author} {\bibinfo {author} {\bibfnamefont {G.}~\bibnamefont
  {Willis}}\ and\ \bibinfo {author} {\bibfnamefont {J.}~\bibnamefont
  {Deardorff}},\ }\bibfield  {title} {\enquote {\bibinfo {title} {Laboratory
  observations of turbulent penetrative-convection planforms},}\ }\href@noop {}
  {\bibfield  {journal} {\bibinfo  {journal} {Journal of Geophysical Research:
  Oceans}\ }\textbf {\bibinfo {volume} {84}},\ \bibinfo {pages} {295--302}
  (\bibinfo {year} {1979})}\BibitemShut {NoStop}%
\bibitem [{\citenamefont {Zdunkowski}\ and\ \citenamefont
  {Nielsen}(1969)}]{zdunkowski}%
  \BibitemOpen
  \bibfield  {author} {\bibinfo {author} {\bibfnamefont {W.~G.}\ \bibnamefont
  {Zdunkowski}}\ and\ \bibinfo {author} {\bibfnamefont {B.~C.}\ \bibnamefont
  {Nielsen}},\ }\bibfield  {title} {\enquote {\bibinfo {title} {A preliminary
  prediction analysis of radiation fog},}\ }\href@noop {} {\bibfield  {journal}
  {\bibinfo  {journal} {pure and applied geophysics}\ }\textbf {\bibinfo
  {volume} {75}},\ \bibinfo {pages} {278--299} (\bibinfo {year}
  {1969})}\BibitemShut {NoStop}%
\bibitem [{\citenamefont {Ponnulakshmi}\ \emph {et~al.}(2012)\citenamefont
  {Ponnulakshmi}, \citenamefont {Mukund}, \citenamefont {~}, \citenamefont
  {Sreenivas},\ and\ \citenamefont {Subramanian}}]{ponnu}%
  \BibitemOpen
  \bibfield  {author} {\bibinfo {author} {\bibfnamefont {V.}~\bibnamefont
  {Ponnulakshmi}}, \bibinfo {author} {\bibfnamefont {V.}~\bibnamefont
  {Mukund}}, \bibinfo {author} {\bibfnamefont {D.}~\bibnamefont {~}}, \bibinfo
  {author} {\bibfnamefont {K.}~\bibnamefont {Sreenivas}}, \ and\ \bibinfo
  {author} {\bibfnamefont {G.}~\bibnamefont {Subramanian}},\ }\bibfield
  {title} {\enquote {\bibinfo {title} {Hypercooling in the nocturnal boundary
  layer: Broadband emissivity schemes},}\ }\href@noop {} {\bibfield  {journal}
  {\bibinfo  {journal} {Journal of Atmospheric Sciences}\ }\textbf {\bibinfo
  {volume} {69}},\ \bibinfo {pages} {2892--2905} (\bibinfo {year}
  {2012})}\BibitemShut {NoStop}%
\bibitem [{\citenamefont {Larson}(2000)}]{larson2000}%
  \BibitemOpen
  \bibfield  {author} {\bibinfo {author} {\bibfnamefont {V.~E.}\ \bibnamefont
  {Larson}},\ }\bibfield  {title} {\enquote {\bibinfo {title} {Stability
  properties of and scaling laws for a dry radiative-convective atmosphere},}\
  }\href@noop {} {\bibfield  {journal} {\bibinfo  {journal} {Quarterly Journal
  of the Royal Meteorological Society}\ }\textbf {\bibinfo {volume} {126}},\
  \bibinfo {pages} {145--171} (\bibinfo {year} {2000})}\BibitemShut {NoStop}%
\bibitem [{\citenamefont {Larson}(2001)}]{larson2001}%
  \BibitemOpen
  \bibfield  {author} {\bibinfo {author} {\bibfnamefont {V.~E.}\ \bibnamefont
  {Larson}},\ }\bibfield  {title} {\enquote {\bibinfo {title} {The effects of
  thermal radiation on dry convective instability},}\ }\href@noop {} {\bibfield
   {journal} {\bibinfo  {journal} {Dynamics of atmospheres and oceans}\
  }\textbf {\bibinfo {volume} {34}},\ \bibinfo {pages} {45--71} (\bibinfo
  {year} {2001})}\BibitemShut {NoStop}%
\end{thebibliography}%

\end{document}